\documentclass[a4paper,11pt]{article}
\pdfoutput=1 % if your are submitting a pdflatex (i.e. if you have
\usepackage{jheppub} % for details on the use of the package, please
\usepackage[T1]{fontenc} % if needed

\usepackage{booktabs}
\usepackage{nicefrac}							%schraege Brueche
\usepackage{textcomp}							%\textcelsius...
\usepackage{upgreek}							%aufrechte griechische Symbole \upmu \updelta...
\usepackage{mathrsfs}							%\mathscr
\usepackage{slashed}							%Feynman slash notation \slashed{...}
\usepackage[utf8]{inputenc}       %Umlaute
\usepackage[german,english]{babel}        %Deutsche Bezeichnungen (zB Inhaltverzeichnis)
\usepackage{amssymb}							%verschiedene Mathematische Symbole
\usepackage{amsmath}							%Formelsatz
\usepackage{dsfont}               %\mathds{...} zB 1 mit Doppelstrich
\usepackage{graphicx}							% This is needed for including figures and graphics
\usepackage[usenames,dvipsnames,svgnames,table]{xcolor}							%generell Farben
\usepackage{float}								%Position von Bildern oder Tabellen
\newcommand{\im}{\mathrm{i}}
\newcommand{\D}{\mathscr{D}}
\newcommand{\halb}{\tfrac{1}{2}}
\newcommand{\ihalbe}{\tfrac{\mathrm{i}}{2}}

\title{Complete superspace classification of three-dimensional Chern-Simons-matter theories coupled  to supergravity}

 \author{Frederik Lauf}
 \author{and Ivo Sachs}
 \affiliation{Ludwig-Maximilians-Universität München,\\Theresienstraße 37, 80333 München, Germany}

\emailAdd{frederik.lauf@physik.lmu.de}
\emailAdd{ivo.sachs@physik.lmu.de}

\abstract{For extended $\mathcal{N}\leq 8$ supersymmetry we classify all possible gauge groups for a scalar multiplet allowed by the algebras of global and local supersymmetry in three dimensions. A detailed discussion of supersymmetry enhancement is included. For the corresponding topologically massive gravity with negative cosmological constant the mass of the graviton is determined algebraically as a function of $\mathcal{N}$ and the possible gauge couplings.}

\begin{document} 
\maketitle
\flushbottom
%\tableofcontents
\newpage
\section*{Introduction and summary}
\addcontentsline{toc}{section}{Introduction and summary}
Superconformal Chern-Simons theories play an important role as conformal field theories describing aspects of M2-branes in string theory, as was suggested in \cite{Schwarz:2004yj} and became clear with the constructions of the BLG \cite{Bagger:2006sk,Gustavsson:2007vu,Bagger:2007jr,Bagger:2007vi} and ABJM \cite{Aharony:2008ug,Aharony:2008gk} models with gauge groups $\mathrm{SO}(4)$ and  $\mathrm{U}(M)\times\mathrm{U}(N)$, inheriting $\mathcal{N}=8$ and $\mathcal{N}=6$ superconformal symmetry  respectively. This motivated the construction and classification of other Chern-Simons-matter theories for various gauge groups, displaying  a certain amount of supersymmetry or vice versa. The first of these models were the $\mathrm{U}(M)\times\mathrm{U}(N)$ and $\mathrm{Sp}(M)\times\mathrm{O}(N)$ theories with $\mathcal{N}=4$ supersymmetry of \cite{Gaiotto:2008sd}. Upon further examination they were found to have also enhanced $\mathcal{N}=5$ and, for $\mathrm{Sp}(M)\times\mathrm{O}(2)$, $\mathcal{N}=6$ supersymmetry in \cite{Hosomichi:2008jb}. A neat group theoretical classification of gauge groups leading to $\mathcal{N}=6$ supersymmetry followed in \cite{Schnabl:2008wj}. The general superconformal gaugings with $\mathcal{N}\leq 8$ where then derived in \cite{Bergshoeff:2008bh} by an approach starting from gauged supergravity. Subsequently, these findings were explained as constraints on possible gauge groups from manifest supersymmetry through superspace approaches for $\mathcal{N}=6$ and $8$ in \cite{Samtleben:2010eu},\cite{Samtleben:2009ts} and \cite{Buchbinder:2008vi}, and similarly for $\mathcal{N}=4$ and $5$ in \cite{Kuzenko:2015lfa} and \cite{Kuzenko:2016cmf}. In this work we will reproduce these results for all cases with $\mathcal{N}\leq 8$ by the superspace method initially employed for $\mathcal{N}=6$ and $8$ in \cite{Gran:2012mg}. This approach, which is universal for all $\mathcal{N}$, focuses on the scalar multiplet transforming under $\mathrm{spin}(\mathcal{N})$ and determining the corresponding on-shell field strengths. The analysis therefore deals mainly with properties of the respective spin matrices which can be shared or bequeathed between different values of $\mathcal{N}$. The simplicity of this method allows direct insight into the mechanism of the supersymmetry enhancement noticed in some of the aforementioned references.

An interesting generalisation is the coupling of these theories to superconformal gravity. This was achieved in \cite{Chu:2009gi} for $\mathcal{N}=6$ (ABJM) and for $\mathcal{N}=8$ in \cite{Gran:2012mg}, the latter giving rise to a new theory with $\mathrm{SO}(N)$ gauge symmetry existing only in the presence of the supergravity sector. The superspace point of view was enabled by the invention \cite{Howe:1995zm} and elaboration \cite{Kuzenko:2011xg} of three dimensional $\mathcal{N}$-extended curved superspace and subsequently by the construction of off-shell actions by the formulation of conformal superspace \cite{Butter:2013goa,Butter:2013rba}. Extending the results for $\mathcal{N}=6$ and $8$ in \cite{Gran:2012mg} we will analyse the constraints for possible gauge groups in curved superspace for $4\leq\mathcal{N}\leq 8$. This will lead to some new models for $\mathcal{N}=6,7$ and $8$, in particular. The following table summarises these findings.
\begin{table}[H]
	\centering
	\begin{tabular*}{\textwidth}{@{\extracolsep{\fill} } lll }
		\toprule
		& fundamental & bifundamental \\\toprule
		$\boldsymbol{\mathcal{N}=4}$ & $\mathrm{SU}(N)_a\times\mathrm{U}(1)_{a-a/N}$ & $\mathrm{U}(M)_a\times\mathrm{U}(N)_{-a}$ \\
		& $\mathrm{Sp}(N)_a\times\mathrm{U}(1)_{-a}$ &  $\mathrm{SU}(M)_a\times\mathrm{SU}(N)_{-a}\times \mathrm{U}(1)_{\nicefrac{a}{N}-\nicefrac{a}{M}}$\\
		& &  $\mathrm{Sp}(M)_a\times\mathrm{SO}(N)_{-a}$\\
		& & $\mathrm{spin}(7)_a\times\mathrm{SU}(2)_{-a}$ \\\midrule
		+SG & & \\
		\midrule
		$\boldsymbol{\mathcal{N}=5}$ & $\mathrm{SU}(N)_a\times\mathrm{U}(1)_{a-a/N}$& $\mathrm{U}(M)_a\times\mathrm{U}(N)_{-a}$  \\
		& $\mathrm{Sp}(N)_a\times\mathrm{U}(1)_{-a}$ &  $\mathrm{SU}(M)_a\times\mathrm{SU}(N)_{-a}\times \mathrm{U}(1)_{\nicefrac{a}{N}-\nicefrac{a}{M}}$\\
		&  &  $\mathrm{Sp}(M)_a\times\mathrm{SO}(N)_{-a}$\\
		& & $\mathrm{spin}(7)_a\times\mathrm{SU}(2)_{-a}$ \\\midrule
		+SG &  & \\
		\midrule
		$\boldsymbol{\mathcal{N}=6}$ & $\mathrm{SU}(N)_a\times\mathrm{U}(1)_{a-a/N}$ & $\mathrm{U}(M)_a\times\mathrm{U}(N)_a$ \\
		& $\mathrm{Sp}(N)_a\times\mathrm{U}(1)_{-a}$ & $\mathrm{SU}(M)_a\times\mathrm{SU}(N)_{a}\times \mathrm{U}(1)_{\nicefrac{a}{N}-\nicefrac{a}{M}}$ \\
		& & $\mathrm{Sp}(M)_a\times\mathrm{SO}(2)_{a}$\\\midrule
		+SG & $\mathrm{SU}(N)\times\mathrm{U}(1)$ & $\mathrm{SU}(M)_a\times\mathrm{SU}(N)_a$ \\
		\midrule
		$\boldsymbol{\mathcal{N}=7}$ & & $\mathrm{SU}(2)_a\times\mathrm{SU}(2)_a$\\\midrule
		+SG & $\mathrm{SU}(N)_{-\nicefrac{\lambda}{8}}\times\mathrm{U}(1)_{(2-N)\nicefrac{\lambda}{16}}$ & $\mathrm{SU}(2)_a\times\mathrm{SU}(2)_{a-\nicefrac{\lambda}{8}}$\\
		&$\mathrm{SO}(N)_{-\nicefrac{\lambda}{16}}$& \\
		& $\mathrm{spin}(7)_{-\nicefrac{\lambda}{16}}$ & \\
		\midrule
		$\boldsymbol{\mathcal{N}=8}$ & & $\mathrm{SU}(2)_a\times\mathrm{SU}(2)_a$\\\midrule
		+SG & $\mathrm{SU}(N)_{-\nicefrac{\lambda}{4}}\times\mathrm{U}(1)_{(2-N)\nicefrac{\lambda}{8}}$& $\mathrm{SU}(2)_a\times\mathrm{SU}(2)_{a-\nicefrac{\lambda}{4}}$\\
		& $\mathrm{SO}(N)_{-\nicefrac{\lambda}{8}}$ &\\
		& $\mathrm{spin}(7)_{-\nicefrac{\lambda}{8}}$ & \\
		\bottomrule
	\end{tabular*}
	\caption{Allowed gauge groups for the ($\mathrm{spin}(\mathcal{N})$-chiral) scalar compensator in flat space and additional groups in the presence of supergravity. Subscripts of the group factors indicate the relative coupling constants (or restricted charges), where for $\mathcal{N}\geq 6$ the Chern-Simons currents corresponding to the right-acting factors will be coupled with opposite sign in the action. Some of the groups with fundamental representation correspond to limits of bifundamental gaugings.}
\end{table}
Focusing on the situation in flat superspace we find a collection of admissible gauge groups for $\mathcal{N}=4$ constituting, in a sequence of increasing $\mathcal{N}$, the first occurrence of a restriction on possible gauge symmetries for fundamental and bifundamental matter (we note in passing that the $\mathrm{spin}(7)$ also implies $\mathrm{G}_2$ gauging, as the subgroup preserving some fixed compensator.) The same restriction appears for $\mathcal{N}=5$. This can be understood by noticing that an $\mathcal{N}=4$ Clifford representation naturally exhibits the same properties as the chiral one, since the left- and right-handed components transform under different factors of $\mathrm{spin}(4)=\mathrm{SU}(2)\times\mathrm{SU}(2)$, while on the other hand, the $\mathcal{N}=4$ Clifford representation just coincides with an implementation of the $\mathcal{N}=5$ spin group $\mathrm{USp}(4)$.
Moving to $\mathcal{N}=6$, many properties of the $\mathcal{N}=5$ matrices as the chiral blocks are taken along; however, the spin group $\mathrm{SU}(4)$ now being manifestly complex prevents the gauging of $\mathrm{SO}(M)\times\mathrm{Sp}(N)$ and $\mathrm{spin}(7)\times\mathrm{SU}(2)$ which rely on a reality condition for the matter fields possible in the previous two cases.
The nature of an $\mathcal{N}=6$ Clifford spinor is quite different from that of an $\mathcal{N}=4$ one. Its left- and right-handed components transform under the $\mathrm{SU}(4)$ representations complex conjugate to each other and appear together in the description in terms of a chiral spinor, whereas for $\mathcal{N}=4$ the two chiral components formed two separate theories. For this reason the gauge groups of the $\mathcal{N}=6$ chiral matter cannot be expected also to be present for a Clifford spinor. It rather turns out that the only possibility is real $\mathrm{SU}(2)\times\mathrm{SU}(2)$ for a Majorana spinor whose chiral components are complex conjugate to each other. This is then the requirement for enhancement to $\mathcal{N}=7$. Finally, this realisation of real $\mathrm{SU}(2)\times\mathrm{SU}(2)$ is transferred to $\mathcal{N}=8$ where the $\mathcal{N}=7$ spin matrices serve as the chiral blocks. 

The additional gauge groups appearing in curved superspace do not show a clear pattern, since the super Cotton tensor which now contributes is of varying rank leading to a different behaviour for each $\mathcal{N}$. The liberation of the $\mathrm{U}(1)$ charges in $\mathcal{N}=6$ is accounted for by the presence of the $\mathrm{U}(1)$ R-symmetry factor in supergravity. The similarity  between $\mathcal{N}=7$ and $8$ is supported by the existence of real and orthogonal representations of their spin groups. In this conformal case, the gravitationally coupled $\mathcal{N}=7$ and $8$ theories are indeed different (although they admit the same gauge groups), which may be worth pointing out in view of results for Poincar\'e supergravity (e.g. \cite{deWit:1992psp}).\\
 
An interesting application of the gravitationally coupled theories is the realisation of topologically massive gravity (TMG) \cite{Deser:1982sw,Chu:2009gi,Gran:2012mg,Nilsson:2013fya,Kuzenko:2013uya, Lauf:2016sac, Kuzenko:2016tfz}. It was noted in \cite{Chu:2009gi} for $\mathcal{N}=6$ (ABJM) that the product of the Chern-Simons coupling and the anti-de Sitter radius is fixed to be $\mu\ell=1$, the chiral point of \cite{Li:2008dq}. For $\mathcal{N}=8$ with $\mathrm{SO}(N)$ gauge symmetry it was found in \cite{Gran:2012mg,Nilsson:2013fya} that $|\mu\ell|^{-1}=3$ for one non-vanishing  component of the compensator and a formula for $p$ components where some of the values correspond, perhaps accidentally, to various interesting TMG solutions. The conjecture that $\mu\ell$ is always fixed in such a superconformal description with $\mathcal{N}\geq 4$ was confirmed in \cite{Lauf:2016sac}. The method there is algebraic and relies on the formalism of $\mathcal{N}$-extended superspace \cite{Howe:1995zm,Kuzenko:2011xg} and conformal superspace \cite{Butter:2013goa,Butter:2013rba} and was used to show that $\mu\ell=1$ for $\mathcal{N}=4$. It will be employed here to determine the values of $\mu\ell$ for $4\leq\mathcal{N}\leq 8$ and all possible deformations due to the presence of additional gauge degrees of freedom and couplings. The result agrees with and extends the values obtained so far in the literature.\\

In section \ref{s1} we give the preliminaries for curved superspaces, scalar multiplets, topologically massive gravity and Chern-Simons gauging. In the subsequent sections we analyse each model with $\mathcal{N}$-extended supergravity, leading to the results outlined above, as well as obtaining on-shell equations for the gauge and supergravity sectors.

\newpage

\section{Superconformal geometry and scalar compensators}\label{s1}
While Chern-Simons-matter theories can be viewed as genuine rigid supersymmetric theories on their own right it is sometime convenient to interpret  the matter fields as compensators for the superconformal geometry. One benefit of this is that one can use superconformal calculus as we do below. Another feature is that it straightforwardly leads to topologically massive supergravities. In this section, as a preparation for the classification of models in the remaining sections, we review the superconformal approach.

\subsection{Conformal superspace and anti-de Sitter superspace}
$\mathcal{N}$-extended superconformal geometry can be formulated in terms of $\mathcal{N}$-extended superspace \cite{Howe:1995zm,Kuzenko:2011xg} (see also\cite{Gran:2012mg}), a curved supermanifold with locally gauged Lorentz and $\mathrm{SO}(\mathcal{N})$ R-symmetry involving super-Weyl invariant constraints on the torsions, or conformal superspace \cite{Butter:2013rba}, where the whole superconformal algebra is gauged as the starting point.

The first formalism, extensively applied here, was used in \cite{Kuzenko:2012bc} to describe anti-de Sitter superspace, where the Lorentz vector fields and the covariant derivatives of the scalar fields forming the torsions are assumed to vanish. In this case, the algebra  of spinor derivatives reduces to
\begin{equation}\label{alg}
	\{\D_\alpha^I,\D_\beta^J\}=2\im\updelta
	^{IJ}(\gamma^a)_{\alpha\beta}\D_a+\im\upvarepsilon_{\alpha\beta}\left(W^{IJKL}+4\updelta^{K[I}K^{J]L}\right)\mathscr{N}_{KL}+4\im K^{IJ}\mathscr{M}_{\alpha\beta}
\end{equation}
where $\mathscr{N}_{KL}$ and $\mathscr{M}_{\alpha\beta}$ denote the R-symmetry and Lorentz generators respectively. The totally antisymmetric tensor $W^{IJKL}$ is the super Cotton multiplet contributing for $\mathcal{N}\geq 4$ and $K^{IJ}=K[\mathrm{diag}(1,...,1,-1,...,-1)]^{IJ}\equiv Kk^{IJ}$ belongs to the compensating Weyl multiplet. If we denote the number of the negative entries by $q$, this is referred to as a $(p,q)$ adS superspace. Furthermore, the commutator of two vector derivatives can be related to the cosmological constant as\footnote{The right-hand side follows from expressing the 3d Riemann tensor by the Ricci tensor which is determined, on shell, by the cosmological constant via the Einstein equation.}
\begin{equation}
	[\D_m,\D_n]|=4K^2|\mathscr{M}_{mn}=\ell^{-2}\mathscr{M}_{mn}.
\end{equation}

In conformal superspace, on the other hand, the algebra of covariant spinor derivatives reads
\begin{equation}
	\{\nabla_\alpha^I,\nabla_\beta^J\}=2\im\updelta
	^{IJ}(\gamma^a)_{\alpha\beta}\nabla_a+\im\upvarepsilon_{\alpha\beta}W^{IJKL}\mathscr{N}_{KL},
\end{equation}
where the special conformal curvatures have been omitted, since those and all higher-dimensional field strengths are expressed by derivatives of the super Cotton tensor $W^{IJKL}$. In the present analysis this formalism will only be used to identify the physical dimension-two $\mathrm{SO}(\mathcal{N})$ field strength, 
\begin{equation}
	F_{(\alpha\beta)}^{IJ}=\tfrac{-\im}{(\mathcal{N}-2)(\mathcal{N}-3)}\nabla_{(\alpha}^K\nabla_{\beta)}^LW^{IJKL}|.\label{fs}
\end{equation}
Restricted to adS superspace, this formula applies to the geometry described by $\D_\alpha^I$, because the extra gauge fields are assumed to vanish.

\subsection{Scalar on-shell multiplets and equations of motion}
The matter fields transform under $\mathrm{spin}(\mathcal{N})$, whose generators are subject to the Lie algebra
\begin{equation}
[\mathscr{N}^{IJ},\mathscr{N}^{KL}]=4\updelta^{[K[I}\mathscr{N}^{J]L]}.
\end{equation}
The Clifford algebra of spin matrices
\begin{equation}
\upgamma^I\upgamma^J=\updelta^{IJ}+\upgamma^{IJ}
\end{equation}
provides a solution as
\begin{align}
\mathscr{N}^{IJ}=-\halb\upgamma^{IJ}.
\end{align}
From the point of view of irreducible representations, the most natural realisation of the spin matrices $\upgamma^{I}$ for $\mathcal{N}=2m$ and $\mathcal{N}=2m+1$ is given by the chiral representation, which can be constructed iteratively as
\begin{align}
	\upgamma^1&=\sigma_1\otimes\mathds{1}\otimes...\otimes\mathds{1}\nonumber\\
	\upgamma^{2,...,2m}&=\im\sigma_2\otimes\im\tilde{\upgamma}^{1,...,\mathcal{N}-1}\nonumber\\
	\upgamma^*&=\upgamma^{2m+1}=-\im^m\upgamma^1\cdot...\cdot\upgamma^{2m}=\sigma_3\otimes\mathds{1}\otimes...\otimes\mathds{1}
\end{align}
where $\tilde{\gamma}$ generate the $\mathcal{N}=2m-1$ dimensional Clifford algebra and each element consists of $m$ factors of $2\times2$ matrices. For even $\mathcal{N}=2m$, the generators of $\mathrm{spin}(\mathcal{N})$ are block-diagonal and commute with the matrices
\begin{align}
	P_{\nicefrac{\mathrm{L}}{\mathrm{R}}}=\halb(\mathds{1}\pm\upgamma^*)
\end{align}
which are projectors on the irreducible representations. These are called left- and right-handed and transform under the generators provided by the chiral Clifford algebra
\begin{align}
	\mathit{\Sigma}^{I}\mathit{\bar{\Sigma}}^{J}=\updelta^{IJ}+\mathit{\Sigma}^{IJ}\nonumber\\
	\mathit{\bar{\Sigma}}^{I}\mathit{\Sigma}^{J}=\updelta^{IJ}+\mathit{\bar{\Sigma}}^{IJ}
\end{align}
where
\begin{equation}
	\upgamma^I=\begin{pmatrix}0&\mathit{\Sigma}^{I}\\\mathit{\bar{\Sigma}}^{I}&0\end{pmatrix}.
\end{equation}
In this case, fields transforming under the chiral generators will be referred to as chiral spinors, whereas those transforming under the reducible generators will be called Clifford spinors.\\

\noindent The algebra (\ref{alg}) acting on a Lorentz scalar transforming under $\mathrm{spin}(\mathcal{N})$ reads\footnote{In the following, $\mathit{\Sigma}^{I}$ is written for spin matrices and may be replaced by $\mathit{\bar{\Sigma}^I}$ or $\upgamma^{I}$ where appropriate.}
\begin{equation}\label{ddq}
	\{\D_\alpha^I,\D_\beta^J\}Q=2\im\updelta
	^{IJ}(\gamma^a)_{\alpha\beta}\D_aQ-\ihalbe\upvarepsilon_{\alpha\beta}\left(W^{IJKL}\mathit{\Sigma}_{KL}+4K^{L[J}\mathit{\Sigma^{I]L}}\right)Q.
\end{equation}
In terms of on-shell superfields, the spinor derivative of the scalar is \cite{Gran:2012mg,Lauf:2016sac}
\begin{equation}
	\D_\alpha^IQ=\im\mathit{\Sigma}^I\mathit{\Lambda}_\alpha.
\end{equation}
In order to obey the supersymmetry algebra, the derivative of $\mathit{\Lambda}_\alpha$ must then be of the form
\begin{equation}
	\D_\alpha^I\mathit{\Lambda}_\beta=(\gamma^a)_{\alpha\beta}\mathit{\Sigma}^I\D_aQ+\halb\upvarepsilon_{\alpha\beta}H^I\label{DLambda}
\end{equation}
where $H^I$ is subject to the equation
\begin{equation}\label{ESG}
	\mathit{\Sigma}^{[J}H^{I]}=-\halb\left(W^{IJKL}\mathit{\Sigma}_{KL}+4K^{L[J}\mathit{\Sigma}^{I]L}\right)Q.
\end{equation}
The general ansatz for $H^I$ is
\begin{equation}\label{Hi}
	H^I=AW^{IKLM}\mathit{\Sigma}_{KLM}Q+BW_{KLPQ}\mathit{\Sigma}^{IKLPQ}Q+2K^{IJ}\mathit{\Sigma}_JQ
\end{equation}
where $A,B$ are constants to be determined using properties of the spin matrices. As will be shown in the corresponding sections, for $\mathcal{N}\leq 6$ this is possible with the supergravity sector being off shell. For $\mathcal{N}=7$ and $8$ it is only possible if the super Cotton tensor is expressed on shell by matter fields (and further, if these are flavoured a solution only exists only in the presence of the corresponding gauge sector).

Being equipped with a solution for $H^I$, the equation of motion for the spinor field $\mathit{\Lambda}_\alpha$ can be obtained by closing the supersymmetry algebra. From the parametrisation (\ref{DLambda}) it follows
\begin{align}
\{\D_\alpha^I,\D_\beta^J\}\mathit{\Lambda_\gamma}=&\mathit{\Sigma}^{(I}\D_{(\alpha}^{J)}\D_{\beta)\gamma}Q-\mathit{\Sigma}^{[I}\D_{[\alpha}^{J]}\D_{\beta]\gamma}Q-\halb\upvarepsilon_{\gamma(\alpha}\D_{\beta)}^{(I}H^{J)}+\halb\upvarepsilon_{\gamma[\alpha}\D_{\beta]}^{[I}H^{J]}.
\end{align}
Commuting the derivatives and keeping only the scalar torsion one finds, using $[\D_\alpha^I,\D_{\beta\gamma}]=-2\upvarepsilon_{\alpha(\beta}\upvarepsilon_{\gamma)\delta}K^{IJ}\D_J^\delta$ \cite{Kuzenko:2012bc},
\begin{align}
	\{\D_\alpha^I,\D_\beta^J\}\mathit{\Lambda_\gamma}=&2\im\updelta^{IJ}\D_{\alpha\beta}\mathit{\Lambda}_\gamma-2\im\updelta^{IJ}\upvarepsilon_{\gamma(\alpha}\slashed{\D}\mathit{\Lambda}_{\beta)}-\im\upvarepsilon_{\alpha\beta}\mathit{\Sigma}^{IJ}\slashed{\D}\mathit{\Lambda}_\gamma\nonumber\\
	&-2\im K^{L(I}\mathit{\Sigma}^{J)}\mathit{\Sigma}^{L}\upvarepsilon_{\gamma(\alpha}\mathit{\Lambda}_{\beta)}+3\im\upvarepsilon_{\alpha\beta}K^{L[I}\mathit{\Sigma}^{J]}\mathit{\Sigma}^{L}\mathit{\Lambda}_\gamma\nonumber\\
	&-\upvarepsilon_{\gamma(\alpha}\D_{\beta)}^{(I}H^{J)}-\halb\upvarepsilon_{\alpha\beta}\D_\gamma^{[I}H^{J]}.
\end{align}
Then it can be read off (with, schematically, $\tilde{H}Q=H$)
\begin{equation}
\slashed{\D}\mathit{\Lambda}_\gamma=-\tfrac{1}{\mathcal{N}}(K_I^{\phantom{I}I}+\halb \tilde{H}_I\mathit{\Sigma}^{I})\mathit{\Lambda}_\gamma.
\end{equation}
Acting with $\D_\beta^J$, antisymmetrising in $\beta\gamma$, and discarding non-scalar background fields, finally leads to the scalar equation of motion
%\begin{align*}
%-\halb\mathit{\Sigma}^{J}\D^{\alpha\beta}\D_{\alpha\beta}Q-K^{JL}H_L=-\tfrac{1}{2\mathcal{N}}(K_I^{\phantom{I}I}+\halb \tilde{H}_I\mathit{\Sigma}^{I})H^J.
%\end{align*}
%Finally it can be written
\begin{align}
\mathcal{N}\D^{a}\D_{a}Q=K^{JL}\mathit{\Sigma}_{J}H_L-\tfrac{1}{2\mathcal{N}}\mathit{\Sigma}_{J}(K_I^{\phantom{I}I}+\halb \tilde{H}_I\mathit{\Sigma}^{I})H^J.
\end{align}

\subsection{Topologically massive gravity}
The action of topologically massive gravity with cosmological constant reads \cite{Deser:1981wh}
\begin{align}
S=-\frac{1}{\kappa^2}\int{\mathrm{d}^3x~e~\left(R+2\ell^{-2}\right)}+\frac{1}{4\mu\kappa^2}\int{\mathrm{d}^3x~e~\upvarepsilon^{mnl}\left(\omega_m^{ab}R_{nl,ab}-\tfrac{2}{3}\omega_m^{ab}\omega_{n,b}^{\phantom{n,b}c}\omega_{l,ca}\right)}.
\end{align}
The superconformal generalisation of this action involves the gravitinos, auxiliary components from the super Cotton tensor and a Chern-Simons term for the $\mathrm{SO}(\mathcal{N})$ gauge fields which is given by \cite{Butter:2013rba}
\begin{equation}
\frac{1}{4\mu\kappa^2}\int{\mathrm{d}^3x~e~\upvarepsilon^{mnl}\left(-2B_m^{IJ}F_{nl,IJ}-\tfrac{4}{3}B_m^{IJ}B_{n,I}^{\phantom{n,I}K}B_{l,KJ}\right)}.\label{cs}
\end{equation}
The Einstein-Hilbert term of the TMG action can be realised by a conformal compensator $\phi$ with the action
\begin{equation}
S=\int{\mathrm{d}^3x~e~\left(-\halb(\overline{\D^a\phi})(\D_a\phi)-\tfrac{1}{16}R|\phi|^2+\lambda(|\phi|^2)^3\right)}\label{cc}
\end{equation}
where the derivatives are covariant with respect to $\mathrm{SO}(\mathcal{N})$ and the trivially acting Lorentz group (and possibly other gauge groups)
\begin{equation}
\D_a=E_a+\halb B_a^{IJ}\mathscr{N}_{IJ}+\halb\mathit{\Omega}_a^{mn}\mathscr{M}_{mn}+... .
\end{equation}
The correct Einstein-Hilbert term is produced if the compensator $\phi$ is chosen such that
\begin{equation}
|\phi|^2=16\kappa^{-2},
\end{equation}
which is possible by a Weyl transformation.

With this coupled compensator, the super Cotton tensor obtains a non-trivial on-shell equation of motion. It can be assumed that
\begin{equation}
W^{IJKL}=c\lambda\bar{Q}\mathit{\Sigma^{IJKL}}Q,
\end{equation}
where $\lambda=\mu\kappa^2$ is the Chern-Simons coupling constant and $c$ is a combinatorial number fixed below. This form is the only possible due to the dimensions of the available fields and---indeed---calculating the field strength with the formula (\ref{fs}) yields
%\footnote{The $\mathcal{N}$-dependent factor cancels due to the identity $\mathit{\Sigma_{KL}}\mathit{\Sigma^{IJKL}}=-(\mathcal{N}-2)(\mathcal{N}-3)\mathit{\Sigma^{IJ}}$.} 
the form of a scalar current
\begin{equation}
F_{(\alpha\beta)}^{IJ}=-c\lambda\left[(\overline{\D_{(\alpha\beta)}\phi})\mathit{\Sigma^{IJ}}\phi-\bar{\phi}\mathit{\Sigma^{IJ}}(\D_{(\alpha\beta)}\phi)\right]
\end{equation}
where $\phi$ is the leading component of $Q$. On the other hand, the scalar current can be read off from the kinetic term in the conformal compensator action (\ref{cc}), while the field strength is related to this current via the equation of motion for the gauge fields from (\ref{cs})
\begin{equation}
-\tfrac{1}{\lambda}\upvarepsilon^{abc}F_{ab}=-\tfrac{2}{\lambda}F^{c}=j^c.
\end{equation}
This determines $c$, leading to the conclusion
\begin{equation}\label{wqq}
W^{IJKL}=-\tfrac{\lambda}{16}\bar{Q}\mathit{\Sigma^{IJKL}}Q.
\end{equation}

With the above results it is possible to determine $\mu\ell$. Imposing $Q$ to be constant, the supersymmetry algebra requires the super-Weyl gauge\footnote{As shown in \cite{Kuzenko:2012bc}, the super Cotton tensor can be non-vanishing only in the case of $(\mathcal{N},0)$ adS superspace, i.e. $k^{IJ}=\updelta^{IJ}$.}
\begin{equation}
	4K\mathit{\Sigma}^{IJ}Q=-W^{IJKL}\mathit{\Sigma}_{KL}Q,
\end{equation}
where we note that an equivalent condition is $H^I=0$. Solving for $K$ and proceeding as in \cite{Lauf:2016sac} one finds 
\begin{equation}
	|\mu \ell|^{-1}Q=\tfrac{1}{2\mathcal{N}(\mathcal{N}-1)}|Q|^{-2}(\bar{Q}\mathit{\Sigma}^{IJKL}Q)\mathit{\Sigma}_{IJKL}Q.\label{formula}
\end{equation}
The value of $|\mu\ell|^{-1}$ is now expressed for general $\mathcal{N}$ in terms of the Fierz identities for the rank-four Clifford matrices. The result is
\begin{center}
	\begin{tabular}{rccccc}
		\toprule
		$\mathcal{N}=$ & 4 & 5 & 6 & 7 & 8\\
		$|\mu\ell|^{-1}=$ & 1 & $\nicefrac{3}{5}$ & 1 & 2 & 3\\
		\bottomrule
	\end{tabular}
\end{center}
where it must be noted that for $\mathcal{N}=6$ the formula had to be adjusted due to an additional $\mathrm{U}(1)_{\mathrm{R}}$ R-symmetry factor without which the theory would not be consistent as will be explained in section 6.

\subsection{Gauge theory}
Gauging a flavour group of the form $F\times G$ with the scalar $Q$ transforming in the bifundamental representation produces a right-acting and a left-acting field strength term subject to the constraint (see e.g. \cite{Kuzenko:2011xg} and references therein)
\begin{equation}
\{\D_\alpha^I,\D_\beta^J\}Q=2\im\updelta^{IJ}(\gamma^a)_{\alpha\beta}\D_aQ+\im\upvarepsilon_{\alpha\beta}F^{IJ}Q+\im\upvarepsilon_{\alpha\beta}QG^{IJ}
\end{equation}
and obeying the Bianchi identity
\begin{equation}
	\D_\alpha^{I}F^{JK}=\D_\alpha^{[I}F^{JK]}-\tfrac{2}{\mathcal{N}-1}\updelta^{I[J}\D_{\alpha,L}F^{K]L}.
\end{equation}
The equation for $H^I$ in the spinor derivative of $\mathit{\Lambda}_\beta$ is now
\begin{equation}\label{cond}
\mathit{\Sigma}^{[J}H^{I]}=F^{IJ}Q+QG^{IJ}.
\end{equation}
The condition for accordance of a gauge group with supersymmetry is tantamount to the existence of a non-zero solution for $H^I$ of this equation. The ansatz
\begin{equation}\label{HI}
H^I=AF^{IK}\mathit{\Sigma}_{K}Q+BF_{KL}\mathit{\Sigma}^{IKL}Q+C\mathit{\Sigma}_{K}QG^{IK}+D\mathit{\Sigma}^{IKL}QG_{KL}
\end{equation}
generally cannot be solved in this off-shell form; however, the field strengths can be specified regarding their algebraic properties by using their on-shell equations \cite{Gran:2012mg}. Given the dimensions of the available fields, these must be rank-two bilinears of the scalars. This agrees with the Bianchi identity and the multiplet projection on the physical dimension-two field strength
\begin{equation}
F_{(\alpha\beta)}\propto\D_{(\alpha}^I\D_{\beta)}^JF_{IJ}\propto\D_{(\alpha}^I\D_{\beta)}^J\bar{Q}\mathit{\Sigma}_{IJ}Q,
\end{equation}
which has the form of a scalar current. The right- and left-acting field strength terms are expressed in terms of the scalars as
\begin{align}
F^{IJ}_A(\tau^A\cdot Q)_{r}^{\phantom{r}\bar{r}}&=a\,\mathrm{tr}(Q\mathit{\Sigma}^{IJ}\tau_A\bar{Q})(\tau^A\cdot Q)_{r}^{\phantom{r}\bar{r}}=a\,Q_v^{\phantom{v}\bar{v}}\mathit{\Sigma}^{IJ}(\tau_A)_w^{\phantom{w}v}\bar{Q}_{\bar{v}}^{\phantom{\bar{v}}w}(\tau^A)_r^{\phantom{r}s}Q_s^{\phantom{s}\bar{r}}\nonumber\\
(Q\cdot\sigma^A)_{r}^{\phantom{r}\bar{r}}G^{IJ}_A&=b\,(Q\cdot\sigma^A)_{r}^{\phantom{r}\bar{r}}\mathrm{tr}(\bar{Q}\mathit{\bar{\Sigma}}^{IJ}\sigma_AQ)=b\,Q_r^{\phantom{r}\bar{s}}(\sigma^A)_{\bar{s}}^{\phantom{\bar{s}}\bar{r}}\bar{Q}_{\bar{v}}^{\phantom{\bar{v}}v}\mathit{\bar{\Sigma}}^{IJ}(\sigma_A)_{\bar{w}}^{\phantom{\bar{w}}\bar{v}}Q_v^{\phantom{v}\bar{w}}
\end{align}
where $a,b$ are the coupling constants and $\tau_A,\sigma_A$ are the generators of the right- and left-acting group factor, respectively. We note that the convenient ordering of $Q$ and $\bar{Q}$ in the right-acting term is opposite to the usual ordering in the kinetic term for the coupled scalar. Therefore, the Chern-Simons current obtains a relative minus sign for $\mathcal{N}=6,7$ and $8$ where the bilinears are antisymmetric as for example $Q\mathit{\Sigma}^{IJ}\bar{Q}=-\bar{Q}\mathit{\bar{\Sigma}^{IJ}}Q$. 

In the case of a fundamental representation we have the field strength term
\begin{equation}
F^{IJ}_A(\tau^A\cdot Q)_r=a\,\bar{Q}^v\mathit{\Sigma}^{IJ}(\tau_A)_v^{\phantom{v}w}Q_w(\tau^A)_r^{\phantom{r}s}Q_s.
\end{equation}
Depending on the group, Fierz-like identities for the generators can be used. The chosen conventions and the resulting field strength terms for the classical gauge groups are presented in the table below (the exceptional cases as in \cite{Bergshoeff:2008bh} will also be considered). Calculating $\mathit{\Sigma}^{[J}H^{I]}$ in terms of these on-shell expressions will reveal the structure of allowed gauge groups.
\begin{table}[H]
	\centering
	\begin{tabular}{ccc}
		\toprule
		 group factor & $(\tau_A)_{ij}(\tau^A)_{kl}$ & $(\bar{Q}\mathit{\Sigma}^{IJ}\tau_AQ)(\tau^AQ)_k$ \\\midrule
		 $\mathrm{SO}(N)$ & $2\updelta_{k[i}\updelta_{j]l}$ & $2(\bar{Q}^{[k}\mathit{\Sigma}^{IJ}Q^{l]})Q_l$\\\addlinespace
		 $\mathrm{Sp}(N)$ & $2\Omega_{k(i}\Omega_{j)l}$ & $2(\bar{Q}^{(k}\mathit{\Sigma}^{IJ}Q^{l)})Q_l$ \\\addlinespace
		 $\mathrm{U}(1)$ & $-q^2\updelta_i^{\phantom{i}j}\updelta_k^{\phantom{k}l}$ & $-q^2(\bar{Q}^l\mathit{\Sigma}^{IJ}Q_l)Q_k$\\\addlinespace
		 $\mathrm{SU}(N)$ & $\tfrac{1}{N}\updelta_i^j\updelta_k^l-\updelta_i^l\updelta_k^j$ & $\tfrac{1}{N}(\bar{Q}^l\mathit{\Sigma}^{IJ}Q_l)Q_k-(\bar{Q}^l\mathit{\Sigma}^{IJ}Q_k)Q_l$ \\\addlinespace
		 $\mathrm{U}(N)$ &  $-\updelta_i^l\updelta_k^j$ & $-(\bar{Q}^l\mathit{\Sigma}^{IJ}Q_k)Q_l$.\\\bottomrule
	\end{tabular}
	\caption{Generator identities and on-shell field strength terms for the classical gauge group factors.}
\end{table}

\section{\boldmath $\mathcal{N}=1$ and $\mathcal{N}=2$}
As a warm-up we discuss the gauging of a scalar multiplet in flat superspace for $\mathcal{N}=2$ and $3$.\footnote{Off-shell Yang-Mills multiplets coupled to conformal supergravity in three spacetime dimensions for $\mathcal{N}\leq 3$ can be found in \cite{Kuzenko:2014jra}.} $\mathcal{N}=1$ gauge theory in three dimensions has been discussed in well-known literature \cite{Gates:1983nr} and we have nothing more to add. 

For $\mathcal{N}=2$, the Clifford algebra is realised by the chiral representation of the spin matrices
\begin{equation}
\upgamma^1=\begin{pmatrix}0&1\\1&0\end{pmatrix}\qquad\upgamma^2=\begin{pmatrix}0&-\im\\\im&0\end{pmatrix}\qquad\upgamma^*=\begin{pmatrix}1&0\\0&-1\end{pmatrix}
\end{equation}
with the chiral blocks
\begin{align}
\mathit{\Sigma}^{1}&=\mathit{\bar{\Sigma}}^{1}=1\nonumber\\
\mathit{\Sigma}^{2}&=-\mathit{\bar{\Sigma}}^{2}=-\im.
\end{align}
The generator of $\mathrm{spin}(2)$ is 
\begin{equation}
\upgamma^{12}=\begin{pmatrix}\im&0\\0&-\im\end{pmatrix}
\end{equation}
and its fundamental representation is reducible into two scalars $Q$ and $\bar{Q}$ transforming under $\mathrm{U}(1)$ and its complex conjugate respectively.

Equation (\ref{cond}) is easily solved for $H^I$ in terms of a complex number
\begin{equation}
\im H^1-H^2=2F^{12}Q+2QG^{12}.
\end{equation}
Any (bi-)fundamental gauging can be implemented in this way.

\section{\boldmath $\mathcal{N}=3$}
The spin matrices are
\begin{align}
\upgamma^1=\begin{pmatrix}0&1\\1&0\end{pmatrix}\qquad\upgamma^2=\begin{pmatrix}0&-\im\\\im&0\end{pmatrix}\qquad\upgamma^3=\begin{pmatrix}1&0\\0&-1\end{pmatrix}
\end{align}
and the generators proportional to $\upgamma^{12},\gamma^{13}$ and $\upgamma^{23}$ are those of $\mathrm{spin}(3)=\mathrm{SU}(2)$. The group indices are raised and lowered as $v^a=\upvarepsilon^{ab}v_b$ and $v_a=v^b\upvarepsilon_{ba}$ where the values of the metric tensor are the entries of
\begin{equation}
\varepsilon=\begin{pmatrix}0&1\\-1&0\end{pmatrix}.
\end{equation}
One can easily show the identity
\begin{equation}\label{ei}
(\upgamma^{[J})_{ab}(\upgamma^{I]})_{cd}=\upvarepsilon_{(a(c}(\upgamma^{IJ})_{d)b)}.
\end{equation}
Rather than substituting on-shell field strengths into (\ref{HI}) it is simpler to directly write down the most general rank one tensor cubic in the Clifford spinor denoted by $q$ and its conjugate $\bar q$
\begin{align}
H^I_a=&(\upgamma^I)_a^{\phantom{a}b}(A\{q_b\bar{q}^cq_c\}+B\{q^c\bar{q}_bq_c\}+C\{q^c\bar{q}_cq_b\})\nonumber\\
&+4(\upgamma^I)_c^{\phantom{c}d}(D\{q^c\bar{q}_dq_a\}+E\{q^c\bar{q}_aq_d\}+F\{q_a\bar{q}^cq_d\}).
\end{align}
The brackets $\{.\}$ encapsulate the group index structure. Since, for the case of a bifundamental representation, there are two free indices, one can have in principle nine terms
\begin{align}
\{A\bar{B}C\}_{r\bar{r}}\equiv\phantom{+}& c_1A_{r\bar{r}}\bar{B}C+c_2A_r\bar{B}_{\bar{r}}C+c_3A_r\bar{B}C_{\phantom{r}\bar{r}}\nonumber\\
+&d_1A_{\phantom{r}\bar{r}}\bar{B}_{\phantom{\bar{r}}r}C+d_2A\bar{B}_{\bar{r}r}C+d_3A\bar{B}_{\phantom{\bar{r}}r}C_{\phantom{r}\bar{r}}\nonumber\\
+&e_1A_{\phantom{r}\bar{r}}\bar{B}C_r+e_2A\bar{B}_{\bar{r}}C_{r}+e_3A\bar{B}C_{r\bar{r}}\,.\label{bifund}
\end{align}
In fact, $d_2$ will always vanish and some of the other terms are usually redundant.\footnote{It must be reminded when the constants $c,d,e$ are implied to be equal in different terms once the constants $A,B,...$ have been related to each other.} The invisible indices are appropriately contracted. For a fundamental representation we define
\begin{equation}
\{A\bar{B}C\}=c_1A_\alpha\bar{B}^\alpha C_\beta+c_2A^\alpha\bar{B}_\beta C_\alpha+c_3A_\beta\bar{B}^\alpha C_\alpha.\label{fund}
\end{equation}
For groups possessing a rank-four invariant, further terms have to be included where the free index is situated at this tensor. This will be relevant for some exceptional groups. Using (\ref{ei}) we then find
\begin{align}
\upgamma^{[J}H^{I]}=&-A(\upgamma^{IJ}\{q)_m\bar{q}^cq_c\}-B\{q^c(\upgamma^{IJ}\bar{q})_mq_c\}-C\{q^c\bar{q}_c(\upgamma^{IJ}q)_m\}\nonumber\\
&+D\{q_m(\bar{q}\upgamma^{IJ}q)+q^c\bar{q}_m(\upgamma^{IJ}q)_c-q^c(\upgamma^{IJ}\bar{q})_mq_c-(\upgamma^{IJ}q)_m\bar{q}^cq_c\}\nonumber\\
&+E\{q_m(\bar{q}\upgamma^{IJ}q)+(q\upgamma^{IJ}\bar{q})q_m-q^c\bar{q}_c(\upgamma^{IJ})_m+(\upgamma^{IJ}q)_m\bar{q}^cq_c\}\nonumber\\
&+F\{q^c\bar{q}_m(\upgamma^{IJ}q)_c+(q\upgamma^{IJ}\bar{q})q_m+q^c\bar{q}_c(\upgamma^{IJ}q)_m+q^c(\upgamma^{IJ}\bar{q})_mq_c\}.
\end{align}
Only the terms which are rank-two bilinears in $q$ and $\bar q$ can contribute to a field strength. The others must cancel out through the choice $D=-F=-B$, $E-D=A$ and $F-E=C$, leading to
\begin{align}
\upgamma^{[J}H^{I]}=&(E+F)\{(q\upgamma^{IJ}\bar{q})q_m\}+(E-F)\{q_m(\bar{q}\upgamma^{IJ}q)\}.
\end{align}
Since the left- and right-acting terms have independent coefficients, any bifundamental or fundamental gauging is possible by choosing the appropriate coefficients in (\ref{bifund}) or (\ref{fund}) according to the table 2. 

We note in passing that upon deleting $\upgamma^3$ in the above equations, the same procedure and result applies to $\mathcal{N}=2$ Clifford spinors.

\section{\boldmath $\mathcal{N}=4$}
This is the minimal number of supersymmetries for which non-trivial constraints on the possible gauge groups as well as the mass of the graviton in toplogically massive gravity are obtained. The left- and right-handed spin matrices are now given by
\begin{align}
(\mathit{\Sigma}^{I})_{i\bar{i}}&=(\mathds{1},\im\sigma_{1,2,3})_{i\bar{i}}\nonumber\\
(\mathit{\bar{\Sigma}}^{I})^{\bar{i}i}&=(\mathds{1},-\im\sigma_{1,2,3})^{\bar{i}i}.
\end{align}
The spin group is $\mathrm{SU}(2)_{\mathrm{L}}\times\mathrm{SU}(2)_{\mathrm{R}}$ where the two factors are associated with the indices $i$ and $\bar{i}$ respectively.
For the rank-four element it holds that
\begin{equation}
\mathit{\Sigma}^{IJKL}=\mathds{1}\upvarepsilon^{IJKL}.
\end{equation}

\subsection{Flavour gauging}
An ansatz for $H^I$ which involves only the left-handed scalar reads
\begin{equation}
H^{I,\bar{m}}=(\mathit{\bar{\Sigma}}^{I})^{\bar{m}m}(A\{Q_m\bar{Q}^iQ_i\}+B\{Q^i\bar{Q}_mQ_i\}+C\{Q^i\bar{Q}_iQ_m\}).
\end{equation}
From this it follows that
%\begin{equation}
%\mathit{\Sigma}^{[J}H^{I]}=(\mathit{\Sigma}^{JI})_k^{\phantom{k}m}(A\{Q_m\bar{Q}^iQ_i\}+B\{Q^i\bar{Q}_mQ_i\}+C\{Q^i\bar{Q}_iQ_m\})
%\end{equation}
\begin{equation}
\mathit{\Sigma}^{[J}H^{I]}=(\mathit{\Sigma}^{JI})^{im}(A\{Q_m\bar{Q}_kQ_i-Q_m\bar{Q}_iQ_k\}+C\{Q_k\bar{Q}_iQ_m-Q_i\bar{Q}_kQ_m\})
\end{equation}
where $B$ has been set to zero without loss of generality and the spinor indices have been rearranged using
\begin{align}
A_iB^kC_k=A^kB_iC_k-A^kB_kC_i
\end{align}
in order to obtain field strength terms.\footnote{This corrects the statement in \cite{Lauf:2016sac} that gauging is not possible with only a one-handed spinor.} The others must be cancelled by choosing $A=C$, leading to
\begin{equation}
\mathit{\Sigma}^{[J}H^{I]}=-A\{(Q\mathit{\Sigma}^{IJ}\bar{Q})Q-Q(\bar{Q}\mathit{\Sigma}^{IJ}Q)\},
\end{equation}
which, in turn, needs to be compatible with (\ref{cond}) for closure of the supersymmetry algebra. 
Applying (\ref{bifund}) we find that a general possibility which avoids field strength terms inconsistent with table 2 is taking only $c_3$ non-zero. In other words one has to consider
\begin{equation}\label{n4c}
\mathit{\Sigma}^{[J}H^{I]}=-Ac_3[(Q\mathit{\Sigma}^{IJ}\bar{Q})Q-Q(\bar{Q}\mathit{\Sigma}^{IJ}Q)],
\end{equation}
where the products are understood as matrix products with the bifundamental indices. Then, the products $\mathrm{U}(M)\times\mathrm{U}(N)$ and $\mathrm{SU}(N)\times\mathrm{SU}(N)$ with opposite couplings $a=-b$ are naturally consistent with (\ref{n4c}), which may be collected into the expression $\mathrm{SU}(M)\times\mathrm{SU}(N)\times\mathrm{U}(1)^\circ$ where the $\mathrm{U}(1)$ charge is constrained by $-q^2=a(\tfrac{1}{N}-\tfrac{1}{M})$ so that it cancels the gauge traced bilinear terms from the $\mathrm{SU}(N)$ factors.\footnote{Clearly, further pairwise cancelling $\mathrm{U}(1)$ factors can always be added.} For these one has $Ac_3=a$, resulting in the solution for $H^I$
\begin{equation}
H^{I,\bar{m}}=a(\mathit{\bar{\Sigma}}^{I})^{\bar{m}m}(Q_m\bar{Q}^iQ_i-Q_i\bar{Q}^iQ_m).
\end{equation}
The only other combination is $\mathrm{Sp}(M)\times\mathrm{SO}(N)$ with opposite couplings and the reality condition
\begin{equation}
Q_i=\bar{Q}^j\upvarepsilon_{ji}.
\end{equation}
This leads to
\begin{equation}
H^{I,\bar{m}}=2a(\mathit{\bar{\Sigma}}^{I})^{\bar{m}m}(Q_m\bar{Q}^iQ_i-Q_i\bar{Q}^iQ_m).
\end{equation}
Another possibility is to consider also a term supplemented to (\ref{bifund}) involving an invariant tensor $C_{ijkl}$. In this case, the second group factor has to be $\mathrm{SU}(2)$ in order to write
\begin{align}
-C_{rvws}Q_v^{\phantom{v}\bar{v}}(\bar{Q}_{\bar{v}w}\mathit{\Sigma}^{IJ}Q_s^{\phantom{s}\bar{r}})=\halb C_{rvws}(Q_v^{\phantom{v}\bar{v}}\mathit{\Sigma}^{IJ}\bar{Q}_{\bar{v}w})Q_s^{\phantom{s}\bar{r}}
\end{align}
where total antisymmetry of $C_{ijkl}$ and reality of $Q$ have been assumed. Since the left-acting symplectic $\mathrm{SU}(2)$ requires the presence of an orthogonal term from the right-acting factor, the generators of the right-acting group must fulfil 
\begin{align}
(\tau^A)_{ij}(\tau_A)_{kl}=2\updelta_{i[k}\updelta_{l]j}+\tfrac{3}{2}C_{ijkl}.
\end{align}
This is the case for $\mathrm{spin}(7)$ (and its subgroup $G_2$) \cite{Bergshoeff:2008bh}. The solution reads
\begin{align}
H^{I,\bar{m}}=&-2a(\mathit{\bar{\Sigma}}^{I})^{\bar{m}m}(Q_m\bar{Q}^iQ_i-Q_i\bar{Q}^iQ_m)_r^{\phantom{r}\bar{r}}\nonumber\\
&-a(\mathit{\bar{\Sigma}}^{I})^{\bar{m}m}C_{rvws}(Q_{m,v}^{\phantom{m,v}\bar{v}}\bar{Q}_{\bar{v}w}Q_s^{\phantom{s}\bar{r}}-Q_{v}^{\phantom{v}\bar{v}}\bar{Q}_{\bar{v}w}Q_{m,s}^{\phantom{m,s}\bar{r}}).
\end{align}
Let us now investigate the possibility of a fundamental representation. In this case one has (setting $c_3=0$ and $A=1$)
\begin{align}
\mathit{\Sigma}^{[J}H^{I]}=\mp c_1\left[(\bar{Q}^\alpha\mathit{\Sigma}^{IJ}Q_\alpha)Q_\beta- (\bar{Q}^\alpha\mathit{\Sigma}^{IJ}Q_\beta)Q_\alpha\right]-(c_2\mp c_2)(\bar{Q}_\beta\mathit{\Sigma}^{IJ}Q^\alpha)Q_\alpha
\end{align}
where the lower sign holds for symplectic groups. Comparing with table 2, we find  that fundamental representations are possible for the groups 
\begin{center}
	\begin{tabular}{cc}
		\toprule
		 $\mathrm{Sp}(N)\times\mathrm{U}(1)^\circ$&$q^2=a$ \\
		 $\mathrm{SU}(N)\times\mathrm{U}(1)^\circ$&$q^2=\tfrac{a}{N}-a$\\
		\bottomrule
	\end{tabular}
\end{center}
with the $\mathrm{U}(1)$ charges being restricted as indicated. These two are equivalent to the above bifundamental $\mathrm{Sp}(N)\times\mathrm{SO}(2)$ and $\mathrm{SU}(N)\times\mathrm{SU}(1)\times\mathrm{U}(1)^\circ$ respectively. For $\mathrm{Sp}(N)\times\mathrm{U}(1)^\circ$ the solution for $H^I$ is
\begin{equation}
H^{I,\bar{m}}=-\tfrac{1}{2}a(\mathit{\bar{\Sigma}}^{I})^{\bar{m}m}\left(Q_{m,\alpha}\bar{Q}^{i,\alpha}Q_{i,\beta}+3Q_{m,\beta}\bar{Q}^{i,\alpha}Q_{i,\alpha}\right)
\end{equation}
and for $\mathrm{SU}(N)\times\mathrm{U}(1)^\circ$
\begin{equation}
H^{I,\bar{m}}=a(\mathit{\bar{\Sigma}}^{I})^{\bar{m}m}\left(Q_{m,\alpha}\bar{Q}^{i,\alpha}Q_{i,\beta}-Q_{i,\alpha}\bar{Q}^{i,\alpha}Q_{m,\beta}\right).
\end{equation}

\subsection{Clifford spinors}
If a right-handed scalar is included in the theory, the ansatz for $H^{I,\bar{m}}$ must be extended by terms like $(\mathit{\bar{\Sigma}^I})^{\bar{m}m}Q_m\bar{Q}_{\bar{i}}Q^{\bar{i}}$ and $(\mathit{\bar{\Sigma}}^{I})^{\bar{k}k}Q_{\bar{k}}\bar{Q}_kQ^{\bar{m}}$. However, the field strength would be proportional to
\begin{equation}
	(Q\mathit{\Sigma}^{IJ}\bar{Q})Q_m+(Q\mathit{\bar{\Sigma}}^{IJ}\bar{Q})Q_m
\end{equation}
for which the conditions producing the left-handed bilinear still would have to apply, leading to the same possible gauge groups. For a more compact description and in prospect of supersymmetry enhancement to $\mathcal{N}=5$, one can use reducible Clifford spinors
\begin{equation}
q_a\stackrel{.}{=}\begin{pmatrix}Q_i\\Q^{\bar{i}}\end{pmatrix}
\end{equation}
with the corresponding spin matrices
\begin{equation}
(\upgamma^{I})_a^{\phantom{a}b}\stackrel{.}{=}\begin{pmatrix}0&(\mathit{\Sigma}^{I})_{i\bar{j}}\\(\mathit{\bar{\Sigma}^I})^{\bar{i}j}&0\end{pmatrix}\quad,\quad(\upgamma^{*})_a^{\phantom{a}b}\stackrel{.}{=}\begin{pmatrix}\updelta_i^{j}&0\\0&-\updelta_{\bar{j}}^{\bar{i}}\end{pmatrix}
\end{equation}
and the metric
\begin{equation}
C^{ab}\stackrel{.}{=}\begin{pmatrix}\upvarepsilon^{ij}&0\\0&\upvarepsilon_{\bar{i}\bar{j}}\end{pmatrix}
\end{equation}
acting by the rules
\begin{align}
q^a=C^{ab}q_b\quad&,\quad q_a=q^bC_{ba}\nonumber\\
Q^i=\upvarepsilon^{ij}Q_j\quad&,\quad Q_i=Q^j\upvarepsilon_{ji}\nonumber\\
Q^{\bar{i}}=\upvarepsilon^{\bar{i}\bar{j}}Q_{\bar{j}}\quad&,\quad Q_{\bar{i}}=Q^{\bar{j}}\upvarepsilon_{\bar{j}\bar{i}}.
\end{align}
We note that, upon including $\upgamma^*$ as $\upgamma^5$, the above realises the spin matrices of $\mathcal{N}=5$. Moreover, the metric $C^{ab}$ coincides with the metric of $\mathrm{spin}(5)=\mathrm{USp}(4)$.

The general ansatz for $H^I_a$ in terms of $q_a$ is then\footnote{Possible terms involving $\upgamma^*$ would turn out to be redundant.}
\begin{align}
	H^I_a=&(\upgamma^{I})_a^{\phantom{a}b}(A\{q_b\bar{q^c}q_c\}+B\{q^c\bar{q}_bq_c\}+C\{q^c\bar{q}_cq_b\})\nonumber\\
	&+2(\upgamma^{I})_c^{\phantom{c}d}(D\{q^c\bar{q}_dq_a\}+E\{q^c\bar{q}_aq_d\}+F\{q_a\bar{q}^cq_d\}).
\end{align}
 For illustration, the $E$-term can be worked out in terms of $\mathrm{SU}(2)$ spinors. Using the identity
\begin{align}
%(\mathit{\Sigma}^{[I})_{i\bar{j}}(\mathit{\Sigma}^{J]})_{k\bar{l}}&=-\halb\upvarepsilon_{ik}(\mathit{\bar{\Sigma}}^{IJ})_{\bar{j}\bar{l}}+\halb\upvarepsilon_{\bar{j}\bar{l}}(\mathit{\Sigma}^{IJ})_{ik}
(\mathit{\Sigma}^{[I})_{i\bar{j}}(\mathit{\bar{\Sigma}}^{J]})^{\bar{l}k}&=-\halb\updelta_i^k(\mathit{\bar{\Sigma}}^{IJ})^{\bar{l}}_{\phantom{\bar{l}}\bar{j}}+\halb\updelta_{\bar{j}}^{\bar{l}}(\mathit{\Sigma}^{IJ})_i^{\phantom{i}k},
\end{align}
it follows that
\begin{align}
	\upgamma^{[J}H'^{I]}%=&2E\{(Q\mathit{\Sigma}^{[I})_{\bar{j}}\begin{pmatrix}(\mathit{\Sigma}^{J]}\bar{Q})_k\\(\mathit{\bar{\Sigma}^{J]}}\bar{Q})^{\bar{k}}\end{pmatrix}Q^{\bar{j}}-(Q\mathit{\bar{\Sigma}^{[I}})^{j}\begin{pmatrix}(\mathit{\Sigma}^{J]}\bar{Q})_k\\(\mathit{\bar{\Sigma}^{J]}}\bar{Q})^{\bar{k}}\end{pmatrix}Q_j\}\nonumber\\
	%=&2E\{\begin{pmatrix}(\mathit{\Sigma}^{[I})_{i\bar{j}}(\mathit{\Sigma}^{J]})_{k\bar{l}}(Q^i\bar{Q}^{\bar{l}}Q^{\bar{j}}-Q^{\bar{j}}\bar{Q}^{\bar{l}}Q^i)\\(\mathit{\Sigma}^{[I})_{i\bar{j}}(\mathit{\bar{\Sigma}}^{J]})^{\bar{k}l}(Q^i\bar{Q}_{l}Q^{\bar{j}}-Q^{\bar{j}}\bar{Q}_{l}Q^i)\end{pmatrix}\}\nonumber\\
	%=&E\{\begin{pmatrix}\left(-\upvarepsilon_{ik}(\mathit{\bar{\Sigma}}^{IJ})_{\bar{j}\bar{l}}+\upvarepsilon_{\bar{j}\bar{l}}(\mathit{\Sigma}^{IJ})_{ik}\right)(Q^i\bar{Q}^{\bar{l}}Q^{\bar{j}}-Q^{\bar{j}}\bar{Q}^{\bar{l}}Q^i)\\\left(-\updelta_i^l(\mathit{\bar{\Sigma}}^{IJ})^{\bar{k}}_{\phantom{\bar{k}}\bar{j}}+\updelta^{\bar{k}}_{\bar{j}}(\mathit{\Sigma}^{IJ})_i^{\phantom{i}l}\right)(Q^i\bar{Q}_{l}Q^{\bar{j}}-Q^{\bar{j}}\bar{Q}_{l}Q^i)\end{pmatrix}\}\nonumber\\
	=&E\{\begin{pmatrix}(\mathit{\Sigma}^{IJ}Q)_k\bar{Q}_{\bar{j}}Q^{\bar{j}}+Q_k(\bar{Q}\mathit{\bar{\Sigma}^{IJ}}Q)+Q_{\bar{j}}\bar{Q}^{\bar{j}}(\mathit{\Sigma}^{IJ}Q)_k-(Q\mathit{\bar{\Sigma}^{IJ}}\bar{Q})Q_k\\(Q\mathit{\Sigma}^{IJ}\bar{Q})Q^{\bar{k}}-Q^j\bar{Q}_j(\mathit{\bar{\Sigma}^{IJ}}Q)^{\bar{k}}-Q^{\bar{k}}(\bar{Q}\mathit{\Sigma}^{IJ}Q)-(\mathit{\bar{\Sigma}^{IJ}}Q)^{\bar{k}}\bar{Q}^jQ_j\end{pmatrix}\}\nonumber\\
	=&E\{(q\upgamma^{IJ}\bar{q})q-q(\bar{q}\upgamma^{IJ}q)-(\upgamma^{IJ}q)\bar{q}^cq_c-q^c\bar{q}_c(\upgamma^{IJ}q)\}.
\end{align}
The second line holds, because terms cubic in $Q$'s of the same handedness cancel identically. Thus, we see that by virtue of their two-component properties bilinears of $Q$ naturally lead to the corresponding bilinears of $q$. The other terms lead to similar structures and must be arranged such that
\begin{align}
\upgamma^{[J}H^{I]}=E\{(q\upgamma^{IJ}\bar{q})q-q(\bar{q}\upgamma^{IJ}q)\},
\end{align}
representing the same situation as for the chiral spinors. The solutions for $H^I$ are now
\begin{center}
	\begin{tabular*}{\textwidth}{@{\extracolsep{\fill} } cc }
		\toprule
		groups & $H^{I,a}$\\\midrule
		$\mathrm{SU}(M)\times\mathrm{SU}(N)\times\mathrm{U}(1)^\circ$ & $a(\upgamma^{I})^{ab}(q_b\bar{q}^cq_c-q_c\bar{q}^cq_b)+2a(\upgamma^{I})^{cd}q_c\bar{q}^aq_d$\\
		$\mathrm{Sp}(M)\times\mathrm{SO}(N)$ & $2a(\upgamma^{I})^{ab}(q_b\bar{q}^cq_c-q_c\bar{q}^cq_b)+4a(\upgamma^{I})^{cd}q_c\bar{q}^aq_d$\\
		$\mathrm{SU}(N)\times\mathrm{U}(1)$ & $a(\upgamma^{I})^{ab}(q_{b}^\alpha\bar{q}^{c}_\alpha q_{c}^\beta-q_{c}^\alpha\bar{q}^{c}_\alpha q_{b}^\beta)+2a(\upgamma^{I})^{cd}q_{c}^\alpha\bar{q}^{a}_\alpha q_{d}^\beta$\\
		$\mathrm{Sp}(N)\times\mathrm{U}(1)$ & $-a(\upgamma^{I})^{ab}(2q_{b}^\alpha\bar{q}^{c}_\alpha q_{c}^\beta-q_{c}^\alpha\bar{q}^{c}_\alpha q_{b}^\beta)-a(\upgamma^{I})^{cd}(2q_{c}^\alpha\bar{q}^{a}_\alpha q_{d}^\beta-q_c^\alpha\bar{q}^{a,\beta}q_{d,\alpha})$.\\
		\bottomrule
	\end{tabular*}
\end{center}
They can be written in the compact form
\begin{align}
H^{I,a}=&-E(\upgamma^{I})^{ab}\{(q_b\bar{q^c}q_c-q_c\bar{q}^cq_b)+2(\upgamma^{I})^{cd}q_c\bar{q}^aq_d\},
\end{align}
with the appropriate coefficients specified above.

In the next section we will recognise that enhancement to $\mathcal{N}=5$ supersymmetry is implicit here, since the above formalism follows trivially from the one for $\mathcal{N}=5$ by removing $\upgamma^5$.

\subsection{Coupling to supergravity}
Referring to (\ref{wqq})  the super Cotton tensor $W^{IJKL}\equiv W\upvarepsilon^{IJKL}$ is given by
\begin{equation}
W=-\tfrac{\lambda}{16}|Q|^2.
\end{equation}
The algebra for pure supergravity (\ref{ddq}) then becomes
\begin{equation}
\{\D_\alpha^I,\D_\beta^J\}Q=2\im\updelta
^{IJ}(\gamma^a)_{\alpha\beta}\D_aQ-\im\upvarepsilon_{\alpha\beta}\left(2K-W\right)\mathit{\Sigma^{IJ}}Q
\end{equation}
and the corresponding solution for $H^I_{\mathrm{\textsc{sg}}}$ is
\begin{equation}
H^I_{\mathrm{\textsc{sg}}}=\left(W-2K\right)\mathit{\bar{\Sigma}^I}Q.
\end{equation}
A constant solution for $Q$ corresponds to $H^I_{\mathrm{\textsc{sg}}}=0$ and leads to $\mu\ell=1$ as shown in \cite{Lauf:2016sac} and as implied by the formula (\ref{formula}).\footnote{For reasons explained in \cite{Lauf:2016sac} the description in terms of a Clifford spinor does not admit  a topologically massive  adS gravity.}

When gauging a flavour symmetry, the super Cotton tensor is expressed as a trace over gauge indices, which clearly is compatible with the above solution for the supergravity sector. The complete solution is then
\begin{equation}
H^I=H^I_{\mathrm{\textsc{sg}}}+H^I_{\mathrm{\textsc{cs}}}
\end{equation}
where $H^I_{\mathrm{\textsc{cs}}}$ is the contribution from the gauge sector of the desired gauge group. This causes a deformation of $\mu\ell$ in terms of the coupling $a$ for the groups with fundamental representation. 

We note that there is no other solution $H^I$ than the above sum, which would both represent the supergravity sector and generalise the gauge groups found in flat superspace.

\section{\boldmath $\mathcal{N}=5$}
Let us begin by recalling the $\mathrm{SO}(5)$ spin matrices $(\upgamma^I)_i^{\phantom{i}j}$ in the chiral representation
\begin{align}
\upgamma_1&=\sigma_1\otimes\mathds{1}\nonumber\\
\upgamma_{2,3,4}&=-\sigma_2\otimes\sigma_{1,2,3}\nonumber\\
\upgamma_5&=\sigma_3\otimes\mathds{1}.
\end{align}
These generate the spin group $\mathrm{USp}(4)$ with invariant symplectic form
\begin{equation}
\varepsilon=\begin{pmatrix}0&1&0&0\\
-1&0&0&0\\
0&0&0&1\\
0&0&-1&0\\\end{pmatrix}.
\end{equation}
Indices are then raised and lowered as
\begin{equation}
q^i=\upvarepsilon^{ij}q_j\quad,\quad q_i=q^j\upvarepsilon_{ji}\quad,\quad\upvarepsilon^{ij}\upvarepsilon_{kj}=\updelta^i_k,
\end{equation}
where $\upvarepsilon_{ij}$ and $\upvarepsilon^{ij}$ are the components of $\varepsilon$. The spin matrices with upper and lower indices are antisymmetric and related by the dualisation
\begin{equation}
\upvarepsilon^{ijkl}(\upgamma^{I})_{kl}=-2(\upgamma^{I})^{ij}.
\end{equation}
This can be used to prove the formula
\begin{equation}
2(\upgamma^{[I})_{ij}(\upgamma^{J]})^{kl}=4\updelta_{[i}^{[k}(\upgamma^{IJ})_{j]}^{\phantom{j}l]}.\label{ranktwo}
\end{equation}
Using antisymmetry and the Clifford-algebra one derives the Fierz identity
%\footnote{Alternative forms are $(\upgamma^I)_{ij}(\upgamma_I)^{kl}=-\upvarepsilon_{ij}\upvarepsilon^{kl}+4\updelta_{[i}^k\updelta_{j]}^l$ and $(\upgamma^I)_{ij}(\upgamma_I)_{kl}=-\upvarepsilon_{ij}\upvarepsilon_{kl}-4\upvarepsilon_{k[i}\upvarepsilon_{j]l}$.}
\begin{align}
(\upgamma^I)_i^{\phantom{i}j}(\upgamma_I)_k^{\phantom{k}l}=-\updelta_i^j\updelta_k^l+2\upvarepsilon_{ik}\upvarepsilon^{jl}+2\updelta_i^l\updelta_k^j.
\end{align}
Finally, the dualisation properties of the rank-two and -four elements are given by
\begin{align}
\upvarepsilon_{I_1...I_5}\upgamma^{I_5}&=\upgamma_{I_1...I_4}\nonumber\\
\upvarepsilon_{I_1...I_5}\upgamma^{I_4I_5}&=-2\upgamma_{I_1...I_3}.
\end{align}

\subsection{Flavour gauging}
For bifundamental scalars, the general ansatz for $H^I_{k}$ is
\begin{align}
H^I_{k}=&(\upgamma^{I})_k^{\phantom{k}l}\left[A\{q_l\bar{q}^iq_i\}+B\{q^i\bar{q}_lq_i\}+C\{q^i\bar{q}_iq_l\}\right]\nonumber\\
&+2(\upgamma^{I})_i^{\phantom{i}j}\left[D\{q^i\bar{q}_jq_k\}+E\{q^i\bar{q}_kq_j\}+F\{q_k\bar{q}^iq_j\}\right].
\end{align}
From this, $\upgamma^{[J}H^{I]}$ can be calculated using (\ref{ranktwo}). Without loss of generality, one can choose the condition $D=F=B=0$ and further $A=C=-E$ in order to cancel all terms which cannot contribute to a field strength. The remaining is
\begin{align}\label{4H}
\upgamma^{[J}H^{I]}=&E\{(q\upgamma^{IJ}\bar{q})q-q(\bar{q}\upgamma^{IJ}q)\}.
\end{align}
Recalling the discussion in section 4 we the see that the possible gauge groups are the same as for $\mathcal{N}=4$. The solution for $H^I$ is then of the form
%There one finds the condition $c_2=-d_2$ (cancelling inappropriate terms) and thus
%\begin{align}
%\upgamma^{[J}H^{I]}=&E\Big(c_1(q\upgamma^{IJ}\bar{q})q+c_3(q^{w\bar{v}}\upgamma^{IJ}\bar{q}_{\bar{v}r})q_w^{\phantom{w}\bar{r}}+d_3(q_r^{\phantom{r}\bar{v}}\upgamma^{IJ}\bar{q}^{\bar{r}w})q_{w\bar{v}}\nonumber\\
%&-c_1q(\bar{q}\upgamma^{IJ}q)-c_3q^{w\bar{v}}(\bar{q}_{\bar{v}r}\upgamma^{IJ}q_w^{\phantom{w}\bar{r}})-d_3q_r^{\phantom{r}\bar{v}}(\bar{q}^{\bar{r}w}\upgamma^{IJ}q_{w\bar{v}})\Big).
%\end{align}
\begin{align}
H^I_k=-E(\upgamma^{I})_k^{\phantom{k}l}\{(q_l\bar{q}^iq_i-q_i\bar{q}^iq_l)+2(\upgamma^{I})^{ij}q_i\bar{q}_kq_j\}.
\end{align}
This clarifies the enhanced $\mathcal{N}=5$ supersymmetry of the $\mathcal{N}=4$ Clifford spinor, for which all of the above equations (and (\ref{ranktwo}), especially,) still hold if we restrict the $\mathrm{SO}(5)$ index to the range  $I=1,...,4$.

\subsection{Coupling to supergravity}
The super Cotton tensor is $W^I=\tfrac{1}{4!}\upvarepsilon^{IJKLM}W_{JKLM}$ and the algebra for pure supergravity becomes
\begin{equation}
\{\D_\alpha^I,\D_\beta^J\}q=2\im\updelta
^{IJ}(\gamma^a)_{\alpha\beta}\D_aq+\im\upvarepsilon_{\alpha\beta}\left(W^A\upgamma_{AIJ}-2K\mathit{\Sigma^{IJ}}\right)q.
\end{equation}
In order to obtain $H^I_{\mathrm{\textsc{sg}}}$, the ansatz (equivalent to (\ref{Hi}))
\begin{equation}
H^I_{\mathrm{\textsc{sg}}}=XW_K\upgamma^{IK}q+YW^Iq+2K\upgamma^Iq
\end{equation}
is inserted into (\ref{ESG}), which yields $X=-Y=-1$.
The on-shell super Cotton tensor is given by
\begin{align}
W^I=-\tfrac{\lambda}{16}|q\upgamma^{I}\bar{q}|,
\end{align}
in terms of which $H^I_{\mathrm{\textsc{sg}}}$ becomes
\begin{align}
H^I_{\mathrm{\textsc{sg}},k}=&-\tfrac{\lambda}{16}(\upgamma^{I}|q)_k\bar{q}^j|q_j+\tfrac{\lambda}{16}|q^j(\upgamma^{I}\bar{q}|)_kq_j\nonumber\\
&+\tfrac{\lambda}{16}|q_k(\bar{q}|\upgamma^{I}q)-\tfrac{\lambda}{16}|q^m\bar{q}_k|(\upgamma^{I}q)_m\nonumber\\
&-\tfrac{\lambda}{16}|q\upgamma^{I}\bar{q}|q_k+2K(\upgamma^Iq)_k,
\end{align}
where $|.|$ denotes the trace over gauge indices.

For the determination of $\mu\ell$, we set $H^I_{\mathrm{\textsc{sg}}}$ on shell to zero and solve for $K$. For this, the contraction $\upgamma_{I}H^I_{\mathrm{\textsc{sg}}}$ has to be evaluated using the Fierz identity. This leads to terms proportional to $\bar{q}^iq_iq_m$ and $q^iq_i\bar{q}_m$, where it is implicitly assumed that $q_i$ is non-vanishing only for one flavour index which can always be achieved by a suitable choice of gauge. The former has the form of the compensator term appearing in $K$, whereas the latter can be handled as follows: Since the metric $\upvarepsilon^{ij}$ consists of block diagonal $2\times2$ $\upvarepsilon$-symbols, one can choose for the compensator $q_{3,4}=0$, leading effectively to a two dimensional object with the metric $\upvarepsilon_{2\times2}$. Then one can use the formula
\begin{align}
A_iB^kC_k=A^kB_iC_k-A^kB_kC_i
\end{align}
valid for such objects to write the problematic terms in the form of the compensator term. With this choice we then find that $|\mu l|^{-1}=\nicefrac{3}{5}.$

In terms of the on-shell super Cotton tensor, the supergravity equation (\ref{ESG}) reads
\begin{align}
\upgamma^{[J}H^{I]}_{\mathrm{\textsc{sg}}}=&\tfrac{\lambda}{16}|q^i\bar{q}_i|(\upgamma^{IJ}q)_k+\tfrac{\lambda}{16}|q_k(\bar{q}|\upgamma^{IJ}q)-\tfrac{\lambda}{16}|q^m\bar{q}_k|(\upgamma^{IJ}q)_m\nonumber\\
&-\tfrac{\lambda}{16}|q^i(\upgamma^{IJ}\bar{q}|)_kq_i+\tfrac{\lambda}{16}(\upgamma^{IJ}|q)_k\bar{q}^l|q_l-2K\upgamma^{IJ}q
\end{align}
which is compatible with the above solution for $H^I_{\mathrm{\textsc{sg}}}$.

Adding the on-shell ansatz for $H^I_{\mathrm{\textsc{sg}}}$ to the ansatz for $H^I_{\mathrm{\textsc{cs}}}$ from the gauge sector, we find no solution that reproduces this equation while generalising the gauge groups already known from the flat case. The gauge sector can lead to deformations of $\mu\ell$ if two gauge components are chosen non-zero for the fundamental matter representations. This behaviour will be treated in the subsequent sections for the theories with higher $\mathcal{N}$, where it will lead to more interesting results.

\section{\boldmath $\mathcal{N}=6$}
This theory is interesting, on one hand due to its relation to M2-branes \cite{Aharony:2008ug,Aharony:2008gk} and also because in this model the coupling to supergravity allows for new flavour gauge groups both in the bifundamental and in the fundamental representation. To see this we first recall the chiral representation of the spin matrices with the adjustment, that the $\mathcal{N}=5$ matrices $\tilde
{\upgamma}^I$ with lower and upper indices are used as the chiral blocks as
\begin{align}
(\upgamma^1)_a^{\phantom{a}b}&=\begin{pmatrix}0&\upvarepsilon_{ij}\\-\upvarepsilon^{ij}&0\end{pmatrix}\equiv\begin{pmatrix}0&(\mathit{\Sigma}^{1})_{ij}\\(\mathit{\bar{\Sigma}}^{1})^{ij}&0\end{pmatrix}\nonumber\\
(\upgamma^{2,...,6})_a^{\phantom{a}b}&=\begin{pmatrix}0&\im(\tilde{\upgamma}^I)_{ij}\\\im(\tilde{\upgamma}^I)^{ij}&0\end{pmatrix}\equiv\begin{pmatrix}0&(\mathit{\Sigma}^{2,...,6})_{ij}\\(\mathit{\bar{\Sigma}}^{2,...,6})^{ij}&0\end{pmatrix}\nonumber\\
(\upgamma^*)_a^{\phantom{a}b}&=\begin{pmatrix}\updelta_i^{\phantom{i}j}&0\\0&-\updelta_i^{\phantom{i}j}\end{pmatrix}.
\end{align}
Basic identities are
\begin{align}
\halb\upvarepsilon^{ijkl}(\mathit{\Sigma}^{I})_{kl}&=-(\mathit{\bar{\Sigma}}^{I})^{ij}\\
(\mathit{\Sigma}^{I})_{ij}(\mathit{\bar{\Sigma}}_{I})^{kl}&=-4\updelta^k_{[i}\updelta^l_{j]}\\
(\mathit{\bar{\Sigma}}^{I})^{ij}(\mathit{\bar{\Sigma}}_{I})^{kl}&=2\upvarepsilon^{ijkl}\\
2(\mathit{\bar{\Sigma}^{[I}})^{ij}(\mathit{\Sigma}^{J]})_{kl}&=4\updelta^{[i}_{[k}(\mathit{\Sigma}^{IJ})_{l]}^{\phantom{l}j]}\label{ie6}
\end{align}
and the dualisation of the rank two element is
\begin{equation}
\upvarepsilon_{IJKLPQ}\mathit{\Sigma}^{PQ}=-2\im\mathit{\Sigma}_{IJKL}.
\end{equation}

\subsection{Flavour gauging}
The general ansatz for $H^I$ is
\begin{equation}\label{6.7}
H^{I,k}=2A(\mathit{\bar{\Sigma}}^{I})^{ij}\{Q_i\bar{Q}^kQ_j\}+B(\mathit{\bar{\Sigma}}^{I})^{kl}\{Q_l\bar{Q}^iQ_i\}+C(\mathit{\bar{\Sigma}}^{I})^{kl}\{Q_i\bar{Q}^iQ_l\}.
\end{equation}
Using (\ref{ie6}) we calculate $\mathit{\Sigma}^{[J}H^{I]}$ in close analogy to $\mathcal{N}=5$ and find the condition $A=B=-C$, so that we are left with\footnote{With this conventional ordering of $Q$ and $\bar{Q}$ the two field strengths in the algebra have the same sign; however, the physical Chern-Simons currents will have opposite signs since $Q\mathit{\bar{\Sigma}}^{IJ}\bar{Q}=-\bar{Q}\mathit{\Sigma}^{IJ}Q$ corresponding to the ordering in the kinetic term.}
\begin{align}
\mathit{\Sigma}^{[J}H^{I]}=&A\{(Q\mathit{\bar{\Sigma}}^{IJ}\bar{Q})Q+Q(\bar{Q}\mathit{\Sigma}^{IJ}Q)\}.
\end{align}
This leads to the solution (cf. \cite{Gran:2012mg})
\begin{align}
H^{I,k}=-A(\mathit{\bar{\Sigma}}^{I})^{kl}(\{Q_l\bar{Q}^iQ_i-Q_i\bar{Q}^iQ_l)+2(\mathit{\bar{\Sigma}}^{I})^{ij}Q_i\bar{Q}^kQ_j\}
\end{align}
known from $\mathcal{N}=4$ and $\mathcal{N}=5$ only this time with the absence of bifundamental $\mathrm{Sp}(M)\times\mathrm{SO}(N>2)$ gauging, since a reality condition is not possible for the complex $\mathrm{SU}(4)$ spinors. This agrees with the classification of \cite{Schnabl:2008wj}. The case of $\mathrm{Sp}(M)\times\mathrm{SO}(2)$ (corresponding to the results of \cite{Hosomichi:2008jb} and \cite{Aharony:2008gk}) is equivalent to the fundamental $\mathrm{Sp}(M)\times\mathrm{U}(1)^\circ$ gauging. Again, restricting the range of $I$ gives an a posteriori demonstration of the supersymmetry enhancement from $\mathcal{N}=4$ to $\mathcal{N}=5$ and (with the mentioned exception) to $\mathcal{N}=6$.

\subsection{Clifford spinors}
In view of a possible enhancement to $\mathcal{N}=7$, we now analyse a possible realisation of these models in terms of a Clifford spinor. We define
\begin{equation}
	q_a=\begin{pmatrix}Q_i\\P^i\end{pmatrix}\quad,\quad\bar{q}^a=\left(\bar{Q}^i,\bar{P}_i\right).
\end{equation}
An ansatz for $H^I$ has to involve terms with $\upgamma^{IK}\upgamma_K$ and $\upgamma^*$, as opposed to $\mathcal{N}=4$ where these terms were superfluous. Working out such an ansatz in terms of the chiral components of the above Clifford spinor, one finds that there remain terms incompatible with the field strengths which cannot be cancelled. In order to move around this obstacle it is necessary to impose the Majorana condition $Q_i=P_i$ and further to assume the bifundamental gauging of real $\mathrm{SU}(2)\times\mathrm{SU}(2)$ where
\begin{equation}
\bar{q}_{\bar{v}}^{\phantom{\bar{v}}v}=\upvarepsilon^{vw}q_w^{\phantom{w}\bar{w}}\upvarepsilon_{\bar{w}\bar{v}}
\end{equation}
so that the identity
\begin{align}
Q_{[k}\bar{Q}_{|i|}Q_{m]}=-\halb Q_{[k}\bar{Q}_{m]}Q_i-\halb Q_i\bar{Q}_{[k}Q_{m]}
\end{align}
applies.\footnote{This follows from $Q_l\mathrm{tr}(\bar{Q}_kQ_m)=Q_k\bar{Q}_mQ_l+Q_m\bar{Q}_kQ_l=Q_l\bar{Q}_kQ_m+Q_l\bar{Q}_mQ_k$.}

However, this is only manifest in a real basis for the Majorana spinor. To this end we perform the transformation
\begin{equation}
q\longrightarrow U\cdot q
\end{equation}
where
\begin{equation}
\begin{pmatrix}Q_i\\\bar{Q}^i\end{pmatrix}\longrightarrow\frac{1}{\sqrt{2}}\begin{pmatrix}\mathds{1}&\mathds{1}\\-\im\mathds{1}&\im\mathds{1}\end{pmatrix}\cdot\begin{pmatrix}Q_i\\\bar{Q}^i\end{pmatrix}=\frac{1}{\sqrt{2}}\begin{pmatrix}Q_i+\bar{Q}^i\\-\im(Q_i-\bar{Q}^i)\end{pmatrix}.
\end{equation}
This defines a real representation since the generators are block-diagonal and it holds that$(\mathit{\Sigma}^{IJ}Q)^*=\mathit{\bar{\Sigma}}^{IJ}\bar{Q}$. Accordingly, the spin matrices are changed to
\begin{align}
\upgamma^I&\longrightarrow U\upgamma^IU^\dagger\nonumber\\
\upgamma^*&\longrightarrow U\upgamma^*U^\dagger.
\end{align}
These are all imaginary and antisymmetric and, with $\upgamma^*=\upgamma^7$, provide a real representation of $\mathrm{spin}(7)$. The index of $q_a$ can now be raised and lowered with the metric $\updelta_{ab}$. 

The Fierz lemma is
\begin{align}
8\updelta_{ab}\updelta_{cd}=&\updelta_{ad}\updelta_{cb}+\upgamma^I_{ad}\upgamma^I_{cb}-\halb\upgamma^{IJ}_{ad}\upgamma^{IJ}_{cb}-\tfrac{1}{3!}\upgamma^{IJK}_{ad}\upgamma^{IJK}_{cb}\nonumber\\
&+\tfrac{1}{4!}\upgamma^{IJKL}_{ad}\upgamma^{IJKL}_{cb}+\tfrac{1}{5!}\upgamma^{IJKLM}_{ad}\upgamma^{IJKLM}_{cb}-\tfrac{1}{6!}\upgamma^{IJKLMN}_{ad}\upgamma^{IJKLMN}_{cb},
\end{align}
which is equivalent to
\begin{align}
8\updelta_{ab}\updelta_{cd}=&\updelta_{(ad)}\updelta_{cb}+\upgamma^I_{[ad]}\upgamma^I_{cb}+\upgamma^{*}_{[ad]}\upgamma^{*}_{cb}-\halb\upgamma^{IJ}_{[ad]}\upgamma^{IJ}_{cb}-\upgamma^{*I}_{[ad]}\upgamma^{*I}_{cb}\nonumber\\
&-\tfrac{1}{3!}\upgamma^{IJK}_{(ad)}\upgamma^{IJK}_{cb}-\tfrac{1}{2}\upgamma^{*IJ}_{(ad)}\upgamma^{*IJ}_{cb}
\end{align}
where the manifest symmetries are indicated. It can be derived
\begin{align}
\upgamma^{I[K}_{ab}\upgamma^{L]I}_{cd}+\upgamma^{*[K}_{ab}\upgamma^{L]*}_{cd}&=4\updelta_{[a[c}\upgamma^{KL}_{d]b]}-\upgamma^{[K}_{ab}\upgamma^{L]}_{cd}\nonumber\\
8\updelta_{(c[b}\upgamma^{K}_{a]d)}&=\updelta_{cd}\upgamma^K_{ab}-\upgamma^{K*J}_{cd}\upgamma^{*J}_{ab}-\halb\upgamma^{KIJ}_{cd}\upgamma^{IJ}_{ab}\nonumber\\
8\updelta_{[c[b}\upgamma^{K}_{a]d]}&=-\upgamma^{KI}_{cd}\upgamma^{I}_{ab}+\upgamma^{KI}_{ab}\upgamma^{I}_{cd}-\upgamma^{K*}_{cd}\upgamma^{*}_{ab}+\upgamma^{K*}_{ab}\upgamma^{*}_{cd}.
\end{align}
As suggested by (\ref{HI}) and the Fierz identities, a more than sufficient ansatz is
\begin{align}
H_b^I=&A(\upgamma^Iq)_b\bar{q}_dq_d+Bq_d(\upgamma^I\bar{q})_bq_d+Cq_d\bar{q}_d(\upgamma^Iq)_b\nonumber\\
&+D(q\upgamma^I\bar{q})q_b+E\upgamma^I_{cd}q_c\bar{q}_bq_d+Fq_b(\bar{q}\upgamma^Iq)\nonumber\\
&+G(q\upgamma^{IK}\bar{q})(\upgamma^Kq)_b+H(\upgamma^Kq)_b(\bar{q}\upgamma^{IK}q)\nonumber\\
&+J(q\upgamma^{I*}\bar{q})(\upgamma^*q)_b+K(\upgamma^*q)_b(\bar{q}\upgamma^{I*}q)\nonumber\\
&+L(q\upgamma^K\upgamma^*\bar{q})(\upgamma^{IK}\upgamma^*q)_b+M(\upgamma^{IK}\upgamma^*q)_b(\bar{q}\upgamma^K\upgamma^*q).
\end{align}
Calculating $\upgamma^{[J}H^{I]}$, most constants can be set to zero while the others are related by $G=-A=B=-D=J$, leading to
\begin{align}
(\upgamma^{[J}H^{I]})_a=G\left[q_a(\bar{q}\upgamma^{IJ}q)+(q\upgamma^{IJ}\bar{q})q_a-\upgamma^{IJ}_{bc}q_b\bar{q}_aq_c)\right].
\end{align}
Using the properties of $\mathrm{SU}(2)\times\mathrm{SU}(2)$, this can be written as
\begin{align}
\upgamma^{[J}H^{I]}=\tfrac{3}{2}G\left[q(\bar{q}\upgamma^{IJ}q)+(q\upgamma^{IJ}\bar{q})q)\right].
\end{align}
The couplings are then $-a=-b=\tfrac{3}{2}G$ and the solution reads
\begin{align}
H_b^I=&\tfrac{2}{3}a\left[(\upgamma^Iq)_b\bar{q}_dq_d-q_d(\upgamma^I\bar{q})_bq_d+(q\upgamma^I\bar{q})q_b-(q\upgamma^{IK}\bar{q})(\upgamma^Kq)_b-(q\upgamma^{I*}\bar{q})(\upgamma^*q)_b\right]
\end{align}
which will be discovered in the next section to be identical to the solution for $\mathcal{N}=7$.

\subsection{Coupling to supergravity}
\noindent The super Cotton tensor is $W^{IJKL}=\halb\upvarepsilon^{IJKLPQ}W_{PQ}$. Since $\mathrm{spin}(6)=\mathrm{SU}(4)$ has no real representation the algebra of the supergravity sector (\ref{ddq}) must be extended to include a $\mathrm{U}(1)_\mathrm{R}$ field strength dual to the super Cotton tensor \cite{Gran:2012mg,Kuzenko:2013vha}, i. e.
\begin{align}
\{\D_\alpha^I,\D_\beta^J\}Q&=2\im(\gamma^a)_{\alpha\beta}\D_aQ-\halb\upvarepsilon_{\alpha\beta}W_{PQ}\mathit{\Sigma}^{IJPQ}Q+\tilde{q}\upvarepsilon_{\alpha\beta}W^{IJ}Q-2\im\upvarepsilon_{\alpha\beta}K\mathit{\Sigma}^{IJ}Q
\end{align}
where $\tilde{q}$ is the $\mathrm{U}(1)_\mathrm{R}$ charge of $Q$. For $H^I_{\mathrm{\textsc{sg}}}$ we find
\begin{equation}
H^I_{\mathrm{\textsc{sg}}}=-\ihalbe W_{PQ}\mathit{\bar{\Sigma}}^{IPQ}Q+\im W^{IK}\mathit{\bar{\Sigma}}_KQ+2K\mathit{\bar{\Sigma}}^IQ
\end{equation}
if $\tilde{q}=-1$. The on-shell super Cotton tensor is
\begin{equation}
W^{IJ}=\tfrac{\lambda}{16}\im|Q\mathit{\bar{\Sigma}}^{IJ}\bar{Q}|,
\end{equation}
so that
\begin{align}
H^{I,k}_{\mathrm{\textsc{sg}}}=-\tfrac{\lambda}{4}|Q_i\bar{Q}^k|(\mathit{\bar{\Sigma}^I})^{il}Q_l+\tfrac{\lambda}{16}|Q_i\bar{Q}^i|(\mathit{\bar{\Sigma}^I})^{kl}Q_l+2K\mathit{\bar{\Sigma}}^IQ.
\end{align}
Then, from $H^I_{\mathrm{\textsc{sg}}}=0$ and one non-vanishing gauge component, it follows $\mu\ell=1$ in agreement with \cite{Chu:2009gi}.

In order to include the gauge sector we recall the condition (\ref{ESG}) from the supergravity sector in terms of scalars
\begin{align}
\mathit{\Sigma}^{[J}H^{I]}_{\mathrm{\textsc{sg}}}=\tfrac{\lambda}{16}\Big[&|Q_i\bar{Q}^i|(\mathit{\Sigma}^{IJ}Q)_k-2(\mathit{\Sigma}^{IJ})_k^{\phantom{k}l}|Q_l\bar{Q}^m|Q_m\nonumber\\
&-2|Q_k\bar{Q}^j|(\mathit{\Sigma}^{IJ}Q)_j+2|Q\mathit{\bar{\Sigma}}^{IJ}\bar{Q}|Q_k\Big]-2K\mathit{\Sigma}^{IJ}Q.
\end{align}
The ansatz  
\begin{equation}
H^I_{\mathrm{\textsc{sg}}}=XW_{PQ}\mathit{\bar{\Sigma}}^{IPQ}Q+YW^{IK}\mathit{\bar{\Sigma}}_{K}Q+2K\mathit{\bar{\Sigma}}^{I}Q
\end{equation}
then gives
\begin{align}
\mathit{\Sigma}^{[J}H^{I]}_{\mathrm{\textsc{sg}}}=&\tfrac{\lambda}{16}[2\im X|Q_i\bar{Q}^i|(\mathit{\Sigma}^{IJ}Q)_k-\im(6X+Y)(\mathit{\Sigma}^{IJ})_k^{\phantom{k}l}|Q_l\bar{Q}^m|Q_m\nonumber\\
&-\im(2X-Y)|Q_k\bar{Q}^j|(\mathit{\Sigma}^{IJ}Q)_j+\im(2X-Y)|Q\mathit{\bar{\Sigma}}^{IJ}\bar{Q}|Q_k]-2K\mathit{\Sigma}^{IJ}Q.
\end{align}
Adding the corresponding ansatz (\ref{6.7}) for $H^I_{\mathrm{\textsc{cs}}}$ of the gauge sector for bifundamental scalars, the system can be solved again. Fixing $X=-\ihalbe$, we find $a=b$, $Y=\im+\im\tfrac{8}{\lambda}a(\tfrac{1}{N}-\tfrac{1}{M})\equiv\im+\im\tfrac{8}{\lambda}af_{NM}$ and
\begin{align}
H^{I,k}_{\mathrm{\textsc{cs}}}=&(\mathit{\bar{\Sigma}}^{I})^{ij}a(f_{NM}|Q_i\bar{Q}^k|Q_j-2Q_i\bar{Q}^kQ_j)\nonumber\\
&+(\mathit{\bar{\Sigma}}^{I})^{kl}a(\tfrac{1}{4}f_{NM}|Q_l\bar{Q}^i|Q_i-\halb f_{NM}Q_l|\bar{Q}^iQ_i|-Q_l\bar{Q}^iQ_i)\nonumber\\
&+a(\mathit{\bar{\Sigma}}^{I})^{kl}Q_i\bar{Q}^iQ_l.
\end{align}
As a result, the coupling to supergravity admits the new possibility $\mathrm{SU}(N)\times\mathrm{SU}(M)$ with equal couplings $a=b$. For $N=M$ these solutions reduce to the $\mathrm{SU}(N)\times\mathrm{SU}(N)$ case with independent gauge and pure supergravity sectors.

Since the compensator term in $K$ is gauge-traced, only one entry of the bifundamental matrix $Q_r^{\phantom{r}\bar{r}}$ can be non-vanishing. In consequence, $\mu\ell$ is deformed in terms of $a$ and $f_{NM}$.\\

For fundamental representations we find $\mathrm{SU}(N)\times\mathrm{U}(1)$ where the $\mathrm{U}(1)$ is arbitrary (as opposed to the flat case) with $X=-\ihalbe$ and $Y=\im-\im\tfrac{8}{\lambda}(\tfrac{a-q^2}{N}-a)$, i.e.
\begin{align}
H_{\mathrm{\textsc{sg}}}^I=&(\tfrac{a-q^2}{N}-a)\bar{Q}^i_\alpha Q_i^\beta(\mathit{\bar{\Sigma}}^{I}Q^\alpha)\nonumber\\
&-(\tfrac{\lambda}{4}-(\tfrac{a-q^2}{N}-a))(Q^\alpha\mathit{\bar{\Sigma}}^{I}Q^\beta)\bar{Q}_\alpha\nonumber\\
&+(\tfrac{\lambda}{16}-\tfrac{1}{2}(\tfrac{a-q^2}{N}-a))\bar{Q}^i_\alpha Q_i^\alpha(\mathit{\bar{\Sigma}}^{I}Q^\beta)+2K\mathit{\bar{\Sigma}}^{I}Q^\beta
\end{align}
and
\begin{align}
H_{\mathrm{\textsc{cs}}}^I=&-(a+\tfrac{a-q^2}{N})(Q^\alpha\mathit{\bar{\Sigma}}^{I}Q^\beta)\bar{Q}^k_\alpha-\tfrac{a-q^2}{N}\bar{Q}^i_\alpha Q_i^\beta(\mathit{\bar{\Sigma}}^{I}Q^\alpha)\nonumber\\
&+\halb(a+\tfrac{a-q^2}{N})(\mathit{\bar{\Sigma}}^{I}Q^\beta)\bar{Q}^k_\alpha Q_k^\alpha.
\end{align}
The sum is
\begin{align}
H^I=H_{\mathrm{\textsc{sg}}}^I+H_{\mathrm{\textsc{cs}}}^I=&(Q^\alpha\mathit{\bar{\Sigma}}^{I}Q^\beta)\bar{Q}^k_\alpha(-2a-\tfrac{\lambda}{4})\nonumber\\
&+\bar{Q}^i_\alpha Q_i^\beta(\mathit{\bar{\Sigma}}^{I}Q^\alpha)(-a)\nonumber\\
&+(\tfrac{\lambda}{16}+a)\bar{Q}^i_\alpha Q_i^\alpha(\mathit{\bar{\Sigma}}^{I}Q^\beta)+2K\mathit{\bar{\Sigma}}^{I}Q^\beta.
\end{align}
This Yang-Mills contribution implies a deformation of $\mu\ell$ in terms of $a$ and $N$. For the particular example $a=-\tfrac{\lambda}{8}$ one has
\begin{align}
H^I=&\tfrac{\lambda}{8}\bar{Q}^i_\alpha Q_i^\beta(\mathit{\bar{\Sigma}}^{I}Q^\alpha)-\tfrac{\lambda}{16}\bar{Q}^i_\alpha Q_i^\alpha(\mathit{\bar{\Sigma}}^{I}Q^\beta)+2K\mathit{\bar{\Sigma}}^{I}Q^\beta.
\end{align}
This can be used to obtain different values for $\mu\ell$ depending on how many of the $\mathrm{SU}(N)$ components are chosen to be non-zero. One can take them as $Q^i_\alpha=v\updelta^i_\alpha$ (as in \cite{Nilsson:2013fya} treating $\mathcal{N}=8$) where $\alpha=1,...,p\leq 4$. This leads to the formula
\begin{equation}
|\mu\ell|^{-1}=\big|\tfrac{2}{p}-1\big|.
\end{equation}

\section{\boldmath $\mathcal{N}=7$}
This model is interesting since it can be coupled to gravity only after gauging the bifundmental flavour symmetry. On the other hand, coupling to gravity leads to new gauge groups in the fundamental representation.  
In the previous section  a real representation of $\mathrm{spin}(7)$ was constructed from a reducible Majorana representation for $\mathcal{N}=6$. Explicitly it is given by
\begin{align}
\gamma^1&=\sigma_1\otimes\mathds{1}\otimes\sigma_2\nonumber\\
\gamma^2&=-\sigma_3\otimes\sigma_1\otimes\sigma_2\nonumber\\
\gamma^3&=-\sigma_1\otimes\sigma_2\otimes\sigma_3\nonumber\\
\gamma^4&=-\sigma_3\otimes\sigma_2\otimes\mathds{1}\nonumber\\
\gamma^5&=-\sigma_1\otimes\sigma_2\otimes\sigma_1\nonumber\\
\gamma^6&=\sigma_3\otimes\sigma_3\otimes\sigma_2\nonumber\\
\gamma^7&=\sigma_2\otimes\mathds{1}\otimes\mathds{1}.
\end{align}
It can be read off
\begin{equation}
(\upgamma^I)_{a(b}(\upgamma^I)_{c)d}=\updelta_{ad}\updelta_{bc}-\updelta_{a(b}\updelta_{c)d}.
\end{equation}
Using the Fierz lemma 
\begin{equation}
8\updelta_{ac}\updelta_{bd}=\updelta_{ad}\updelta_{bc}+\upgamma^{I}_{ad}\upgamma^{I}_{bc}-\halb\upgamma^{IJ}_{ad}\upgamma^{IJ}_{bc}-\tfrac{1}{6}\upgamma^{IJK}_{ad}\upgamma^{IJK}_{bc}
\end{equation}
one can derive the useful identities\footnote{The first one can be used to calculate $\mu\ell$ with the formula given in the beginning.}
\begin{align}
\upgamma^{IJK}_{ab}\upgamma^{IJK}_{cd}&=6(\updelta_{ab}\updelta_{cd}-8\updelta_{c(a}\updelta_{b)d})\nonumber\\
\upgamma^{I[K}_{ab}\upgamma^{L]I}_{cd}&=4\updelta_{[a[c}\upgamma^{KL}_{d]b]}-\upgamma^{[K}_{ab}\upgamma^{L]}_{cd}\nonumber\\
\upgamma^{IJ[K}_{ab}\upgamma^{L]IJ}_{cd}&=-8\updelta_{(c(a}\upgamma^{KL}_{b)d)}\nonumber\\
-8\updelta_{(c[a}\upgamma^{K}_{b]d)}&=\updelta_{cd}\upgamma^K_{ab}-\halb\upgamma^{KIJ}_{cd}\upgamma^{IJ}_{ab}\nonumber\\
-8\updelta_{[c[a}\upgamma^{K}_{b]d]}&=-\upgamma^{KI}_{cd}\upgamma^{I}_{ab}+\upgamma^{KI}_{ab}\upgamma^{I}_{cd}\nonumber\\
%\upgamma^{I}_{ab}\upgamma^{IKL}_{cd}&=\halb\upgamma^{IJ}_{ab}\upgamma^{IJKL}_{cd}-8\updelta_{(c[a}\upgamma^{KL}_{b]d)}-\upgamma^{KL}_{ab}\updelta_{cd}\\
\upgamma^{I}_{ab}\upgamma^{IKL}_{cd}&=-\upgamma^{KL}_{ab}\updelta_{cd}-4\updelta_{(c[a}\upgamma^{KL}_{b]d)}+4\upgamma^{[K}_{(c[a}\upgamma^{L]}_{b]d)}\nonumber\\
\upgamma^{IJ}_{ab}\upgamma^{IJKL}_{cd}&=8\updelta_{(c[a}\upgamma^{KL}_{b]d)}+8\upgamma^{[K}_{(c[a}\upgamma^{L]}_{b]d)}\nonumber\\
\upgamma^{KLM}_{cd}\upgamma^{IKLM}_{ab}&=48\updelta_{(a(c}\upgamma^{I}_{d)b)}
\end{align}
The spin matrices are related by the dualisations
\begin{align}
%\upvarepsilon^{SPQRKLM}\mathit{\Sigma}_{M}&=\im\mathit{\Sigma}^{SPQRKL}\\
\upvarepsilon^{SPQRKLM}\upgamma_{LM}&=2\im\upgamma^{SPQRK}\nonumber\\
\upvarepsilon^{SPQRKLM}\upgamma_{KLM}&=-3!\im\upgamma^{SPQR}.
\end{align}

\subsection{Flavour gauging}
Noting that there is no closed form for the contraction $\upgamma^I_{ab}\upgamma^I_{cd}$ we need to include more terms in the ansatz for $H^I$ than before. As suggested by the off-shell form (\ref{HI}) and the Fierz identities it is sufficient to write
\begin{align}
H_a^I=&\upgamma_{ab}^I\left[A\{q_b\bar{q}_cq_c\}+B\{q_c\bar{q}_bq_c\}+C\{q_c\bar{q}_cq_b\}\right]\nonumber\\
&+\upgamma_{cd}^I\left[D\{q_c\bar{q}_dq_a\}+E\{q_c\bar{q}_aq_d\}+F\{q_a\bar{q}_cq_d\}\right]\nonumber\\
&+\upgamma^{IK}_{cd}\upgamma^{K}_{ab}\left[G\{q_c\bar{q}_dq_b\}+H\{q_b\bar{q}_cq_d\}\right].
\end{align}
Evaluating $\upgamma^{[J}H^{I]}$, one is forced to set $G=-D$, $H=-F$ and $E=0$ in order to cancel the terms involving $\upgamma^{[I}_{ij}\upgamma^{J]}_{kl}$ and further to choose $C=H$, $A=-G$, $G-H=B$. Finally, we take $H=0$ without loss of generality. This leads to
\begin{align}
(\upgamma^{[J}H^{I]})_a=&G\{(q\upgamma^{IJ}\bar{q})q_a+q_a(\bar{q}\upgamma^{IJ}q)-\upgamma^{IJ}_{bc}q_b\bar{q}_aq_c\}.
%&+H\{-\upgamma^{IJ}_{jl}Q_j\bar{Q}_mQ_l+(Q\upgamma^{IJ}\bar{Q})Q_m+Q^m(\bar{Q}\upgamma^{IJ}Q)\}
\end{align}
The last term can only be dealt with in the case of real $\mathrm{SU}(2)\times\mathrm{SU}(2)$ so that 
\begin{align}
\upgamma^{[J}H^{I]}=&\tfrac{3}{2}G\left[(q\upgamma^{IJ}\bar{q})q+q(\bar{q}\upgamma^{IJ}q)\right].
\end{align}
Then one can take $Gc_3=-\tfrac{2}{3}a=-\tfrac{2}{3}b$ resulting in
\begin{align}
H_a^I=&\tfrac{2}{3}a\upgamma_{ab}^I\left[q_b\bar{q}_cq_c-q_c\bar{q}_bq_c\right]+\tfrac{2}{3}a\upgamma_{cd}^Iq_c\bar{q}_dq_a-\tfrac{2}{3}a\upgamma^{IK}_{cd}\upgamma^{K}_{ab}q_c\bar{q}_dq_b
\end{align}
which is the same as for the $\mathcal{N}=6$ Majorana spinor if the value $I=7$ of the free index is excluded and the term in the contraction with $K=7$ is extracted in terms of $\upgamma^*$. There is no solution for a fundamental gauge group.
\subsection{Coupling to supergravity}
The super Cotton tensor is $W^{IJKL}=\tfrac{1}{6}\upvarepsilon^{IJKLSPQ}W_{SPQ}$, so that \begin{equation}
\upgamma^{[J}H^{I]}_{\mathrm{\textsc{sg}}}=-\halb\left(-\tfrac{\im}{3}W_{SPQ}\upgamma^{IJSPQ}+4K\upgamma^{IJ}\right)q.
\end{equation}
The ansatz
\begin{align}
H^I_{\mathrm{\textsc{sg}}}=XW_{KLM}\upgamma^{IKLM}q+YW^{IKL}\upgamma_{KL}q+2K\upgamma^Iq
\end{align}
leads to
\begin{align}
\upgamma^{[J}H^{I]}_{\mathrm{\textsc{sg}}}=-XW_{SPQ}\upgamma^{IJSPQ}q+(Y-3X)W^{PQ[I}\upgamma^{J]PQ}q+2YW^{IJP}\upgamma_{P}q.
\end{align}
This time, there is no solution for $H^I$ in this off-shell form. On shell, with 
\begin{equation}
W^{IJK}=-\im\tfrac{\lambda}{16}|q\upgamma^{IJK}\bar{q}|
\end{equation}
one finds
\begin{align}
-\halb W^{IJKL}\upgamma_{KL}q=&-\tfrac{\lambda}{16}\left[|q(\bar{q}|\upgamma^{IJ}q)+|q_c(\upgamma^{IJ}\bar{q}|)q_c+(\upgamma^{IJ}|q)\bar{q}_c|q_c+|q_c\bar{q}|(\upgamma^{IJ}q)_c\right]\nonumber\\
&+\tfrac{\lambda}{8}\left[(\upgamma^{[I}|q)(\bar{q}|\upgamma^{J]}q)+|q_c(\upgamma^{[I}\bar{q}|)(\upgamma^{J]}q)_c\right]
\end{align}
and for the ansatz
\begin{align}
H^I_{\mathrm{\textsc{sg}}}=&-4\tfrac{-\im\lambda}{16}(Y-3X)\left[|q(\bar{q}|\upgamma^Iq)+|q_c\bar{q}|(\upgamma^Iq)_c\right]\nonumber\\
&-4\tfrac{-\im\lambda}{16}(Y+3X)\left[|q_c(\upgamma^I\bar{q}|)q_c+(\upgamma^I|q)\bar{q}_c|q_c\right]\nonumber\\
&+2Y\tfrac{-\im\lambda}{16}|q_c\bar{q}_c|(\upgamma^Iq)\nonumber\\
&+2K\upgamma^Iq,
\end{align}
implying
\begin{align}
\upgamma^{[J}H^{I]}_{\mathrm{\textsc{sg}}}=&-4\tfrac{-\im\lambda}{16}(Y-3X)\left[|(\upgamma^{[J}q)(\bar{q}|\upgamma^{I]}q)+|q_c(\upgamma^{[J}\bar{q})|(\upgamma^{I]}q)_c\right]\nonumber\\
&+4\tfrac{-\im\lambda}{16}(Y+3X)\left[|q_c(\upgamma^{IJ}\bar{q}|)q_c+(\upgamma^{IJ}|q)\bar{q}_c|q_c\right]\nonumber\\
&-2Y\tfrac{-\im\lambda}{16}|q_c\bar{q}_c|(\upgamma^{IJ}q)\nonumber\\
&-2K\upgamma^{IJ}q.
\end{align}
Also here, the coefficients cannot be chosen to reproduce the supergravity term, except in the absence of flavour indices where there is a solution with $X=0$ and $Y=-\tfrac{\im}{3}$ (which implies $|\mu\ell|^{-1}=2$).

For a flavoured scalar, we add the ansatz $H^I_{\mathrm{\textsc{sg}}}$ for the supergravity sector to the ansatz $H^I_{\mathrm{\textsc{cs}}}$ for the gauge sector and evaluate again $\mathit{\Sigma}^{[J}H^{I]}$. For bifundamental matter it turns out that real $\mathrm{SU}(2)\times\mathrm{SU}(2)$ remains the only possibility. In this case the gauge traces can be turned into matrix products in the bifundamental indices so that
\begin{align}
-\halb W^{IJKL}\upgamma_{KL}q=&-\tfrac{\lambda}{8}\left[\halb q(\bar{q}\upgamma^{IJ}q)+q_c(\upgamma^{IJ}\bar{q})q_c+(\upgamma^{IJ}q)\bar{q}_cq_c-\halb(q\upgamma^{IJ}\bar{q})q\right]\nonumber\\
&+\tfrac{\lambda}{4}\left[(\upgamma^{[I}q)(\bar{q}\upgamma^{J]}q)+q_c(\upgamma^{[I}\bar{q})(\upgamma^{J]}q)_c\right]
\end{align}
and
\begin{align}
\upgamma^{[J}H^{I]}=&-8\tfrac{-\im\lambda}{16}(Y-3X)\left[(\upgamma^{[J}q)(\bar{q}\upgamma^{I]}q)+q_c(\upgamma^{[J}\bar{q})(\upgamma^{I]}q)_c\right]\nonumber\\
&+8\tfrac{-\im\lambda}{16}(Y+3X)\left[q_c(\upgamma^{IJ}\bar{q})q_c+(\upgamma^{IJ}q)\bar{q}_cq_c\right]-4Y\tfrac{-\im\lambda}{16}q_c\bar{q}_c(\upgamma^{IJ}q)\nonumber\\
&-A(\upgamma^{IJ}q)\bar{q}_cq_c+G(\tfrac{3}{2}q(\bar{q}\upgamma^{IJ}q)+\tfrac{3}{2}(q\upgamma^{IJ}\bar{q})q-(\upgamma^{IJ}q)\bar{q}_cq_c)\nonumber\\
&-2K\upgamma^{IJ}q
\end{align}
where we have set some coefficients to zero without loss of generality and further $G=B=-D$. It can be seen that the field strength terms of the gauge sector also have to contribute to the supergravity sector. This leads to the relation for the couplings $a-b=\tfrac{\lambda}{8}$ and $a+b=-3G$. Fixing the remaining constants (e.g. $X=-Y=-\tfrac{\im}{8}$, $A=-\tfrac{\lambda}{32}-G$) finally leads to the solution
\begin{align}
H^I=&\left[\tfrac{\lambda}{8}+\tfrac{1}{3}(2a-\tfrac{\lambda}{8})\right]((\upgamma^Iq)\bar{q}_cq_c+(q\upgamma^I\bar{q})q)+\left[\tfrac{\lambda}{8}-\tfrac{1}{3}(2a-\tfrac{\lambda}{8})\right]q_c(\upgamma^I\bar{q})q_c\nonumber\\
&-\tfrac{\lambda}{8}q(\bar{q}\upgamma^Iq)-\tfrac{1}{3}(2a-\tfrac{\lambda}{8})(q\upgamma^{IK}\bar{q})(\upgamma^Kq)\nonumber\\
&+2K\upgamma^{I}q.
\end{align}
Giving an expectation value to one of the $\mathrm{spin}(7)$ components implies $|\mu\ell|^{-1}=2$. 

Regarding fundamental gauge groups, it is expected that $\mathrm{SU}(N)$ is a possibility at least for $N=2$ since $a$ or $b$ can be set to zero in the above bifundamental gauging. Indeed, for $\mathrm{SU}(N)\times\mathrm{U}(1)$ it is found a solution where the coupling is completely fixed by the gravity coupling $a=-\tfrac{\lambda}{8}$ and the charge is constrained as $q^2=\tfrac{\lambda}{16}(N-2)$. The corresponding solution for $H^I$ reads
\begin{align}
H^I_\beta=&-\tfrac{\lambda}{8}\left[q_\alpha(\bar{q}^\alpha\upgamma^Iq_\beta)+q^c_\alpha\bar{q}^\alpha(\upgamma^Iq_\beta)_c+q^c_\alpha(\upgamma^I\bar{q}^\alpha)q^c_\beta\right]\nonumber\\
&+\tfrac{\lambda}{16}\left[(\upgamma^Iq_\alpha)\bar{q}^\alpha_cq^c_\beta-(\upgamma^Iq_\beta)\bar{q}^\alpha_c q_\alpha^c-(q_\alpha\upgamma^{[I}\bar{q}^\alpha)(\upgamma^{J]}q_\beta)+(q_\alpha\upgamma^{IK}\bar{q}^\alpha)(\upgamma^Kq_\beta)\right]\nonumber\\
&+2K\upgamma^{I}q_\beta.
\end{align}
For $\mathrm{SO}(N)$, a field strength term can be provided entirely by the supergravity sector, with $a=-\tfrac{\lambda}{16}$ and
\begin{align}
H^I_\beta=\tfrac{\lambda}{4}q_\alpha(q_\alpha\upgamma^Iq_\beta)+\tfrac{\lambda}{8}(\upgamma^Iq_\alpha)q_\alpha q_\beta+2K\upgamma^{I}q_\beta.
\end{align}
Then, also $\mathrm{spin}(7)$ or $G_2$ can be gauged by employing the invariant rank-four tensor in the ansatz for the gauge sector leading to the term
\begin{align}
GC_{\beta\alpha\gamma\delta}[(q^\alpha\upgamma^{IJ}q^\gamma)q_a^\delta+q_a^\alpha(q^\gamma\upgamma^{IJ}q^\delta)-\upgamma^{IJ}_{bc}q_b^\alpha q_a^\gamma q_c^\delta]=3GC_{\beta\alpha\gamma\delta}(q^\alpha\upgamma^{IJ}q^\gamma)q_a^\delta.
\end{align}
Then $G=-\tfrac{\lambda}{16}\tfrac{1}{2}$ and
\begin{align}
H^I_\beta=&\tfrac{\lambda}{4}q_\alpha(q_\alpha\upgamma^Iq_\beta)+\tfrac{\lambda}{8}(\upgamma^Iq_\alpha)q_\alpha q_\beta+\tfrac{\lambda}{32}C_{\beta\alpha\gamma\delta}[(q^\alpha\upgamma^Iq^\gamma) q^\delta-(q^\alpha\upgamma^{IK}q^\gamma)\upgamma^{K}q^\delta]\nonumber\\
&+2K\upgamma^{I}q_\beta.
\end{align}

\section{\boldmath $\mathcal{N}=8$}
The spin matrices of $\mathcal{N}=7$ are taken as the chiral blocks $(\mathit{\Sigma}^{I})_{i\bar{i}}$ and $(\mathit{\bar{\Sigma}}^{I})_{\bar{i}i}$ for $\mathcal{N}=8$, namely
\begin{align}
\mathit{\Sigma}^{1}&=\mathit{\bar{\Sigma}}^{1}=\mathds{1}\nonumber\\
\mathit{\Sigma}^{2,...,8}&=-\mathit{\bar{\Sigma}}^{2,...,8}=\im\tilde{\gamma}^{1,...,7}
\end{align}
so that $(\mathit{\Sigma}^{I})^{\mathrm{T}}=\mathit{\bar{\Sigma}}^{I}$. We note the ``triality relation''\footnote{A list of many identities can be found in \cite{Samtleben:2009ts}.}
\begin{equation}
(\mathit{\Sigma}^{I})_{i(\bar{i}}(\mathit{\Sigma}^{I})_{j\bar{j})}=\updelta_{ij}\updelta_{\bar{i}\bar{j}}.
\end{equation}
It indicates that interchanging the role of the $\mathrm{SO}(8)$ indices with that of one of the $\mathrm{spin}(8)$ matrix indices specifies new spin matrices solving the Clifford algebra. For the superspace it is then formally possible to let the spinor coordinates transform under one of the $\mathrm{spin}(8)$ representations while the scalar multiplet carries an $\mathrm{SO}(8)$ vector index. The resulting, algebraically equivalent formalism was used for the BLG model, especially. We will repeat the following treatment in this ``trialised version'' in the appendix.
\subsection{Flavour gauging}
The ansatz for $H^I$ is very similar to the one for $\mathcal{N}=7$ and the relevant Fierz identity needed to calculate $\mathit{\Sigma}^{[J}H^{I]}$ is
\begin{equation}
(\mathit{\Sigma}^{K[I})_{ij}(\mathit{\Sigma}^{J]K})_{kl}=4\updelta_{[i[k}(\mathit{\Sigma}^{IJ})_{l]j]}
\end{equation}
which may be derived by enhancing the $\mathcal{N}=7$ identities to include $\mathit{\Sigma}^{1}=\mathit{\bar{\Sigma}}^1=\mathds{1}$. The general ansatz for $H^I$ is then
\begin{align}
H_{\bar{k}}^I=&(\mathit{\bar{\Sigma}^I})_{\bar{k}k}\left[A\{Q_k\bar{Q}_iQ_i\}+B\{Q_i\bar{Q}_kQ_i\}\right]+C(\mathit{\Sigma}^{IK})_{ij}(\mathit{\bar{\Sigma}^K})_{\bar{k}k}\{Q_i\bar{Q}_jQ_k\}.
\end{align}
It implies
\begin{align}
(\mathit{\Sigma}^{[J}H^{I]})_m=&-\mathit{\Sigma}^{IJ}_{mk}\left[A\{Q_k\bar{Q}_iQ_i\}+B\{Q_i\bar{Q}_kQ_i\}\right]+C\{(Q\mathit{\Sigma}^{IJ}\bar{Q})Q_m+Q_m(\bar{Q}\mathit{\Sigma}^{IJ}Q)\nonumber\\
&+Q_i(\mathit{\Sigma}^{IJ}\bar{Q})_mQ_i-\mathit{\Sigma}^{IJ}_{kl}Q_k\bar{Q}_mQ_l-(\mathit{\Sigma}^{IJ}Q)_m\bar{Q}_iQ_i\}
\end{align}
Again, one finds the sole possibility of real $\mathrm{SU}(2)\times\mathrm{SU}(2)$ and
\begin{align}
H_{\bar{k}}^I=&\tfrac{2}{3}a(\mathit{\bar{\Sigma}^I})_{\bar{k}k}\left[Q_k\bar{Q}_iQ_i-Q_i\bar{Q}_kQ_i\right]-\tfrac{2}{3}a(\mathit{\Sigma}^{IK})_{ij}(\mathit{\bar{\Sigma}^K})_{\bar{k}k}Q_i\bar{Q}_jQ_k.
\end{align}
The equivalence to the solution with explicit $\mathrm{SO}(7)$ covariance is totally obvious. For a fundamental representation there is no solution.

\subsection{Comment on supersymmetry enhancement}
This concludes a line from the $\mathcal{N}=4$ to the $\mathcal{N}=8$ chiral theories. Supersymmetry enhancement in this framework has two aspects. The first one, from the $\mathcal{N}=4$ Clifford theory via $\mathcal{N}=5$ to the $\mathcal{N}=6$ chiral theory and similarly from $\mathcal{N}=6$ Majorana to $\mathcal{N}=8$ chiral, means that one can extend the index range of the $\mathrm{SO}(\mathcal{N})$ vector index $I$ in the supersymmetry transformations
\begin{align}
\D_\alpha^IQ&=\im\mathit{\Sigma}^I\mathit{\Lambda}_\alpha\nonumber\\
\D_\alpha^I\mathit{\Lambda}_\beta&=(\gamma^a)_{\alpha\beta}\mathit{\Sigma}^I\D_aQ+\halb\upvarepsilon_{\alpha\beta}H^I
\end{align}
without changing the form of these equations, i.e. the form of $H^I$ in particular. The second one concerns the critical transition from $\mathcal{N}=4$ chiral to $\mathcal{N}=4$ Clifford and similarly for $\mathcal{N}=6$. Here, it is crucial whether it is possible to construct a Clifford doublet from two chiral spinors while keeping the structure leading to the allowed gauge symmetries. For $\mathcal{N}=4$ this is naturally the case, which is owed to the two-component properties of the chiral spinors and to the fact that both $\mathrm{spin}(4)$ and $\mathrm{spin}(5)$ (i.e. $\mathrm{spin}(4)$ Clifford) are symplectic groups. A different case occurs from $\mathcal{N}=6$ chiral to $\mathcal{N}=6$ Clifford. Due to the complexness of $\mathrm{spin}(6)=\mathrm{SU}(4)$ it turns out that a doublet of two $\mathrm{SU}(4)$ spinors has very different algebraic properties than a single spinor. The only case where one can implement a gauge symmetry is real $\mathrm{SU}(2)\times\mathrm{SU}(2)$ which can be realised by imposing the Majorana condition on the Clifford spinor. This real representation then extends to $\mathcal{N}=7$ and $8$ as discussed above.

\subsection{Coupling to supergravity}
The super Cotton tensor is now self-dual and the supergravity equation continues to be
\begin{align}
\mathit{\Sigma}^{[J}H^{I]}_{\mathrm{\textsc{sg}}}=-\halb W^{IJKL}\mathit{\Sigma}_{KL}Q-2K\mathit{\Sigma}^{IJ}Q.
\end{align}
The ansatz
\begin{equation}
H^I_{\mathrm{\textsc{sg}}}=XW^{IKLM}\mathit{\bar{\Sigma}}_{KLM}Q+2K\mathit{\bar{\Sigma}}^IQ
\end{equation}
provides no solution off shell since
\begin{align}
\mathit{\Sigma}^{[J}H^{I]}_{\mathrm{\textsc{sg}}}&=XW^{[I|KLM|}\mathit{\Sigma}^{J]KLM}Q+3XW^{IJLM}\mathit{\Sigma}_{LM}Q-2K\mathit{\Sigma}^{IJ}Q.
\end{align}
In terms of the on-shell super Cotton tensor
\begin{equation}
W^{IJKL}=-\tfrac{\lambda}{16}|Q\mathit{\Sigma}^{IJKL}\bar{Q}|
\end{equation}
the super Cotton term reads
\begin{align}
-\halb W^{IJKL}\mathit{\Sigma}_{KL}Q%=&\tfrac{\lambda}{32}|Q_i\bar{Q}_j|Q_l\left[2\updelta_{ij}\mathit{\Sigma}^{IJ}_{kl}-16\updelta_{[k(i}\mathit{\Sigma}^{IJ}_{j)l]}\right]\nonumber\\
=&\tfrac{\lambda}{16}|Q_i\bar{Q}_i|(\mathit{\Sigma}^{IJ}Q)-\tfrac{\lambda}{8}|Q_i(\mathit{\Sigma}^{IJ}\bar{Q}|)Q_i-\tfrac{\lambda}{8}(\mathit{\Sigma}^{IJ}|Q)\bar{Q}_i|Q_i\nonumber\\
&-\tfrac{\lambda}{8}|Q(\bar{Q}|\mathit{\Sigma}^{IJ}Q)-\tfrac{\lambda}{8}|Q_i\bar{Q}|(\mathit{\Sigma}^{IJ}Q)_i
\end{align}
while the ansatz becomes
\begin{align}
H^I_{\mathrm{\textsc{sg}}}=&-X\tfrac{\lambda}{16}\left[24|Q_i(\mathit{\bar{\Sigma}^I}\bar{Q}|)Q_i+24(\mathit{\bar{\Sigma}^I}|Q)\bar{Q}_i|Q_i-6|Q_i\bar{Q}_i|(\mathit{\bar{\Sigma}^I}Q)\right]\nonumber\\
&+2K\mathit{\bar{\Sigma}}^IQ
\end{align}
and implies
\begin{align}
\mathit{\Sigma}^{[J}H^{I]}_{\mathrm{\textsc{sg}}}=&X\tfrac{\lambda}{16}\left[24|Q_i(\mathit{\Sigma}^{IJ}\bar{Q}|)Q_i+24(\mathit{\Sigma}^{IJ}|Q)\bar{Q}_i|Q_i-6|Q_i\bar{Q}_i|(\mathit{\Sigma}^{IJ}Q)\right]\nonumber\\
&-2K\mathit{\Sigma}^{IJ}Q.
\end{align}
Assuming an unflavoured scalar leads to $X=\tfrac{1}{14}$ and
\begin{equation}
H^I_{\mathrm{\textsc{sg}}}=3\tfrac{\lambda}{16}Q^2\mathit{\Sigma}^{I}Q+2K\mathit{\Sigma}^{I}Q.
\end{equation}
Taking $H^I_{\mathrm{\textsc{sg}}}=0$ leads to $|\mu\ell|^{-1}=3$.

For a flavoured scalar we again add the on-shell ansatz of the gauge sector to the one for the supergravity sector. It becomes apparent that for bifundamental matter the only possibility remains real $\mathrm{SU}(2)\times\mathrm{SU}(2)$ in which case
\begin{align}
-\halb W^{IJKL}\mathit{\Sigma}_{KL}Q=&\tfrac{\lambda}{8}Q_i\bar{Q}_i(\mathit{\Sigma}^{IJ}Q)-\tfrac{\lambda}{4}Q_i(\mathit{\Sigma}^{IJ}\bar{Q})Q_i-\tfrac{\lambda}{4}(\mathit{\Sigma}^{IJ}Q)\bar{Q}_iQ_i\nonumber\\
&-\tfrac{\lambda}{4}Q(\bar{Q}\mathit{\Sigma}^{IJ}Q)-\tfrac{\lambda}{4}Q_i\bar{Q}(\mathit{\Sigma}^{IJ}Q)_i
\end{align}
and
\begin{align}
\mathit{\Sigma}^{[J}H^{I]}%=&X\tfrac{\lambda}{16}\left[24|Q_i(\mathit{\Sigma}^{IJ}\bar{Q}|)Q_i+24(\mathit{\Sigma}^{IJ}|Q)\bar{Q}_i|Q_i-6|Q_i\bar{Q}_i|(\mathit{\Sigma}^{IJ}Q)\right]\nonumber\\
%&-\mathit{\Sigma}^{IJ}_{mk}\left[A\{Q_k\bar{Q}_iQ_i\}+B\{Q_i\bar{Q}_kQ_i\}\right]+C\{(Q\mathit{\Sigma}^{IJ}\bar{Q})Q_m+Q_m(\bar{Q}\mathit{\Sigma}^{IJ}Q)\nonumber\\
%&+Q_i(\mathit{\Sigma}^{IJ}\bar{Q})_mQ_i-\mathit{\Sigma}^{IJ}_{kl}Q_k\bar{Q}_mQ_l-(\mathit{\Sigma}^{IJ}Q)_m\bar{Q}_iQ_i\}\\
=&X\tfrac{\lambda}{16}\left[48Q_i(\mathit{\Sigma}^{IJ}\bar{Q})Q_i+36(\mathit{\Sigma}^{IJ}Q)\bar{Q}_iQ_i\right]-2K\mathit{\Sigma}^{IJ}Q\nonumber\\
&-\left[A(\mathit{\Sigma}^{IJ}Q)\bar{Q}_iQ_i+BQ_i(\mathit{\Sigma}^{IJ}\bar{Q})Q_i\right]\nonumber\\
&+\tfrac{3}{2}C\left[(Q\mathit{\Sigma}^{IJ}\bar{Q})Q+Q(\bar{Q}\mathit{\Sigma}^{IJ}Q)+Q_i(\mathit{\Sigma}^{IJ}\bar{Q})Q_i-(\mathit{\Sigma}^{IJ}Q)\bar{Q}_iQ_i\right]
\end{align}
where the gauge traces have been rewritten as matrix products in the bifundamental indices. One finds a solution where the couplings must fulfil $a+b=-3C$ and $a-b=\tfrac{\lambda}{4}$ and
%, while e.g. $X=-\tfrac{1}{12}$, $C=B$ and $-A=C+\tfrac{\lambda}{16}$. Then
\begin{align}
H^I=&\tfrac{\lambda}{16}\left[\tfrac{16}{3}Q_i(\mathit{\bar{\Sigma}^I}\bar{Q})Q_i+\tfrac{2}{3}(\mathit{\bar{\Sigma}^I}Q)\bar{Q}_iQ_i\right]+\tfrac{2}{3}a\left[(\mathit{\bar{\Sigma}^I}Q)\bar{Q}_iQ_i-Q_i(\mathit{\bar{\Sigma}^I}\bar{Q})Q_i\right]\nonumber\\
&-\tfrac{1}{3}(2a-\tfrac{\lambda}{4})(\mathit{\Sigma}^{IK})_{ij}Q_i\bar{Q}_j(\mathit{\bar{\Sigma}^K}Q)+2K\mathit{\bar{\Sigma}}^IQ.
\end{align}
This can be rewritten in terms of traces which gives $|\mu\ell|^{-1}=3$ for one non-vanishing $\mathrm{spin}(8)$ component.

For a fundamental representation at least $\mathrm{SU}(2)$ should be possible, as $a$ or $b$ can be set to zero in the above bifundamental gauging. Indeed, more generally for $\mathrm{SU}(N)\times\mathrm{U}(1)$ it is found $a=-\tfrac{\lambda}{4}$ and the $\mathrm{U}(1)$ charge obeys $q^2=(N-2)\tfrac{\lambda}{8}$. %With $X=-\tfrac{1}{6}$ and $A=-2C=-\tfrac{\lambda}{4}$
The solution then is
\begin{align}
H^I_\beta=\tfrac{\lambda}{16}\left[4Q^i_\alpha(\mathit{\bar{\Sigma}^I}\bar{Q}^\alpha)Q^i_\beta-Q^i_\alpha\bar{Q}_i^\alpha(\mathit{\bar{\Sigma}^I}Q_\beta)\right]+\tfrac{\lambda}{8}(Q_\beta\mathit{\Sigma}^{IK}\bar{Q}^\alpha)(\mathit{\bar{\Sigma}^K}Q_\alpha)+2K\mathit{\bar{\Sigma}}^IQ_\beta.
\end{align}
For $\mathrm{SO}(N)$, the field strength term is contained in the supergravity sector with the coupling $a=-\tfrac{\lambda}{8}$. It follows
%the combined ansatz for $H^I$ leads to
%\begin{align}
%\mathit{\Sigma}^{[J}H^{I]}_\beta=&X\tfrac{\lambda}{16}\left[48(\mathit{\Sigma}^{IJ}Q_\alpha)Q^i_\alpha Q^i_\beta-6Q^i_\alpha Q^i_\alpha(\mathit{\Sigma}^{IJ}Q_\beta)\right]-2K\mathit{\bar{\Sigma}}^{IJ}Q_\beta\nonumber\\
%&-A(\mathit{\Sigma}^{IJ}Q_\alpha)Q^i_\alpha Q^i_\beta.
%\end{align}
%A solution is $X=-\tfrac{1}{6}$, $A=-\tfrac{\lambda}{4}$ and the coupling is fixed to be $a=-\tfrac{\lambda}{8}$. It follows
\begin{align}
H^I_\beta=\tfrac{\lambda}{16}\left[4(\mathit{\bar{\Sigma}^I}Q_\alpha)Q_\alpha^iQ^i_\beta-Q_\alpha^iQ_\alpha^i(\mathit{\bar{\Sigma}^I}Q_\beta)\right]+2K\mathit{\bar{\Sigma}}^IQ_\beta.
\end{align}
This leads to the formula already discovered in \cite{Nilsson:2013fya}
\begin{equation}
|\mu\ell|^{-1}=\big|\tfrac{4}{p}-1\big|
\end{equation}
where $p\leq8$ is the number of non-vanishing entries of the matrix $Q_i^\alpha=\mathrm{diag}(v,...,v,0,...,0)$. Finally, for $\mathrm{spin}(7)$ or $G_2$ one has
\begin{align}
H^I_\beta=&\tfrac{\lambda}{16}\left[4(\mathit{\bar{\Sigma}^I}Q_\alpha)Q_\alpha^iQ^i_\beta-Q_\alpha^iQ_\alpha^i(\mathit{\bar{\Sigma}^I}Q_\beta)\right]-\tfrac{\lambda}{16}C_{\beta\alpha\gamma\delta}(\mathit{\Sigma}^{IK})_{ij}(\mathit{\bar{\Sigma}^K})_{\bar{k}k}Q_i^\alpha Q_j^\gamma Q_k^\delta\nonumber\\
&+2K\mathit{\bar{\Sigma}}^IQ_\beta.
\end{align}
\section{Conclusions}
In this paper we have elaborated on the on-shell superspace formulation of Chern-Simons-matter theories with and without coupling to supergravity, introduced in \cite{Gran:2012mg}. The strength of this formalism is that the classification of such theories is to a large extend reduced to representation theory of the $\mathrm{spin}(\mathcal{N})$ R-symmetry group and therefore provides a unifying view on theories with different numbers of supercharges. Moreover, it readily provides the matter equations of motion useful for model building.

While confirming (and correcting some) results in the previous literature and revealing the relation between models with different numbers of supercharges within our construction, we completed the classification of such models by a number of new consistent theories coupled to supergravity, for $\mathcal{N}=6,7$ and $8$ in particular. We hope that some of these models will be useful in string/M-theory. 

We also found a plethora of new topologically massive gravity models with enhanced supersymmetry and determined the masses of the graviton in these theories. Perhaps, this can be a good starting point for analysing the non-perturbative consistency of topologically massive gravity. 

\textbf{Acknowledgements.} This work was supported by the DFG Transregional Collaborative Research Centre TRR 33 and the DFG cluster of excellence ”Origin and Structure of the Universe”.

\appendix
\section{\boldmath $\mathcal{N}=8$ in the trialised version}
\subsection*{Pure supergravity}
The supercoordinates now transform under $\mathrm{spin}(8)$ and the scalar multiplet under $\mathrm{SO}(8)$. Hence,
\begin{equation}
\{\D_\alpha^i,\D_\beta^j\}Q^I=2\im\updelta^{ij}(\gamma^a)_{\alpha\beta}\D_aQ^I+\im\upvarepsilon_{\alpha\beta}W^{ijkl}\mathscr{N}_{kl}Q^I+4\im\upvarepsilon_{\alpha\beta}K\mathscr{N}^{ij}Q^I,
\end{equation}
where 
\begin{equation}
\mathscr{N}_{kl}Q^I=\halb(\mathit{\Sigma}^{IJ})_{kl}Q_J.
\end{equation}
An antisymmetric tensor is converted to a spinorial tensor as\footnote{In agreement with $\mathscr{N}_{KL}Q_I=-2\updelta_{I[K}Q_{L]}$.}
\begin{equation}
-\tfrac{1}{4}(\mathit{\Sigma}^{IJ})_{ij}A_{IJ}=A_{ij}.
\end{equation}
The derivatives of the scalar and spinor fields are
\begin{equation}
\D_{\alpha,i}Q^I=\im(\mathit{\Sigma}^{I})_{i\bar{i}}\mathit{\Lambda}_\alpha^{\bar{i}}
\end{equation}
and
\begin{equation}
\D_\alpha^i\mathit{\Lambda}_\beta^{\bar{j}}=(\gamma^a)_{\alpha\beta}(\mathit{\Sigma}^{I})^{i\bar{j}}\D^aQ_I+\halb\upvarepsilon_{\alpha\beta}H^{i\bar{j}}
\end{equation}
where $H^{i\bar{j}}$ must fulfil
\begin{equation}
(H\mathit{\bar{\Sigma}}^{I})^{[ij]}=\halb W^{ijkl}(\mathit{\Sigma}^{IJ})_{kl}Q_J+2K(\mathit{\Sigma}^{IJ})^{ij}Q_J.
\end{equation}
The self-dual super Cotton tensor can be expressed by a symmetric and traceless rank-two tensor \cite{Gran:2012mg}
\begin{equation}
W_{ijkl}\equiv\tfrac{1}{16}(\mathit{\Sigma}^{KP})_{[ij}(\mathit{\Sigma}^{LP})_{kl]}C_{KL}
\end{equation}
leading to
\begin{align}
(H\mathit{\bar{\Sigma}}^{I})^{[ij]}&=\tfrac{1}{2}C^{K[I}(\mathit{\Sigma}^{J]K})_{ij}Q_J+2K(\mathit{\Sigma}^{IJ})^{ij}Q_J.
\end{align}
The exclusive ansatz for $H_{i\bar{j}}$ is
\begin{equation}
H_{i\bar{j}}=A(\mathit{\Sigma}^{K})_{i\bar{j}}C_{KJ}Q^J-2K(\mathit{\Sigma}^{J})_{i\bar{j}}Q_J
\end{equation}
implying
\begin{align}
(\mathit{\Sigma}^{I}\bar{H})_{ij}&=-AC^{K[I}\mathit{\Sigma}^{J]K}Q_J+AC^{K(I}\mathit{\Sigma}^{J)K}Q_J+2K(\mathit{\Sigma}^{IJ})^{ij}Q_J.
\end{align}
This means that, thanks to the second term, the algebra is not consistent if the super Cotton tensor is off shell. On shell, due to its symmetry and tracelessness, the super Cotton tensor is of the form \cite{Gran:2012mg}
\begin{equation}
C_{IJ}=C\left(\bar{Q}_{(I}Q_{J)}-\tfrac{1}{8}\updelta_{IJ}\bar{Q}^KQ_K\right).
\end{equation}
For the above expressions it is found
\begin{align}
C^{KI}\mathit{\Sigma}^{JK}Q_J=-\halb C\mathit{\Sigma}^{KJ}\left(\bar{Q}^{I}Q^K+\bar{Q}^KQ^{I}\right)Q_J-\tfrac{1}{8}C\mathit{\Sigma}^{JI}\bar{Q}^KQ^KQ_J.
\end{align}
Assuming no flavour gauging so far, one finds the solution $A=-\tfrac{3}{14}$ and
\begin{equation}
H_{i\bar{j}}=-\tfrac{3}{16}C(\mathit{\Sigma}^{K})_{i\bar{j}}Q_{K}Q^JQ_{J}-2K(\mathit{\Sigma}^{J})_{i\bar{j}}Q_J.
\end{equation}
Setting $H_{i\bar{j}}=0$ leads to the relation
\begin{equation}
\ell^{-1}=\tfrac{3}{16}\tilde{C}\lambda Q^JQ_{J}=3\tilde{C}\mu
\end{equation}
where $\tilde{C}\lambda\equiv C$.

In order to specify the constant $C$, the equation of motion for the $\mathrm{SO}(8)$ gauge field has to be determined. The third component of the super Cotton tensor in the trialised version is given by
\begin{equation}
-\halb F_{\alpha\beta}^{ij}=w_{\alpha\beta}^{ij}=\tfrac{\im}{60}\nabla_{(\alpha}^k\nabla_{\beta)}^lW^{ijkl}.
\end{equation}
The spinor super Cotton tensor is on-shell expressed by $C^{IJ}$
\begin{equation}
W^{ijkl}=\tfrac{C}{16}(\mathit{\Sigma}^{IK})^{[ij}(\mathit{\Sigma}^{JK})^{kl]}Q_{(I}Q_{J)}.
\end{equation}
It can be calculated
\begin{equation}
\nabla_{(\alpha}^k\nabla_{\beta)}^lQ_{(I}Q_{J)}=2\im(\gamma^a)_{\alpha\beta}(\mathit{\Sigma}^{K(I})_{kl}Q^{J)}\nabla_aQ_K
\end{equation}
leading to
\begin{align}
w_{\alpha\beta}^{ij}=\tfrac{-1}{180}\tfrac{C}{16}(\gamma^a)_{\alpha\beta}\left[32(\mathit{\Sigma}^{IL})^{ij}\updelta^{JL}_{K(I}Q_{J)}+4(\mathit{\Sigma}^{IL}\mathit{\Sigma}^{K(I}\mathit{\Sigma}^{|JL|})^{[ij]}Q^{J)}\right]\nabla_aQ_K
\end{align}
and after further elaboration it can be found
\begin{equation}
w_{\alpha\beta}^{ij}=\tfrac{C}{16}(\gamma^a)_{\alpha\beta}(\mathit{\Sigma}^{JK})^{ij}Q_{J}\nabla_aQ_K.
\end{equation}
Comparing the kinetic term (omitting the term quadratic in $B_a$)
\begin{align}
-\halb\D^aQ^I\D_aQ_I=\tfrac{1}{4}B_a^{ij}(\mathit{\Sigma}^{IJ})_{ij}Q_I\partial_aQ_J\equiv -j^a_{ij}B_a^{ij}
\end{align}
with the equation of motion
\begin{equation}
\tfrac{2}{\lambda}F_a^{ij}=j_a^{ij}
\end{equation}
yields $C=\tilde{C}\lambda=\lambda$ and thus $|\mu\ell|^{-1}=3$.
\subsection*{Flavour gauging}
The supersymmetry algebra for a gauge group $F\times G$ is
\begin{equation}
\{\D_\alpha^i,\D_\beta^j\}Q^K=\im\updelta^{ij}\D_{\alpha\beta}Q^K+\im\upvarepsilon_{\alpha\beta}F^{ij}Q^K+\im\upvarepsilon_{\alpha\beta}Q^KG^{ij}.
\end{equation}
The condition for $H$ reads
\begin{equation}
(\mathit{\Sigma}^{I}\bar{H})_{[ij]}=F_{ij}Q^I+Q^IG_{ij}=a(\mathit{\Sigma}^{KL})_{ij}\{Q_K\bar{Q}_LQ^I\}+b(\mathit{\Sigma}^{KL})_{ij}\{Q^I\bar{Q}_KQ_L\}.
\end{equation}
The ansatz is\footnote{An off-shell ansatz cannot be solved.}
\begin{align}
\bar{H}=&A\mathit{\bar{\Sigma}}^{KLM}\{Q_K\bar{Q}_LQ_M\}\nonumber\\
&+\mathit{\bar{\Sigma}}^{K}(B\{Q_J\bar{Q}_KQ_J\}+C\{Q_K\bar{Q}_JQ_J\}+D\{Q_J\bar{Q}_JQ_K\}).
\end{align}
It is found immediately $B=C=D=0$ and thus
\begin{align}
\mathit{\Sigma}^{I}\bar{H}=&A(\mathit{\Sigma}^{LM}\{Q_I\bar{Q}_LQ_M\}+\mathit{\Sigma}^{KL}\{Q_K\bar{Q}_LQ_I\}-\mathit{\Sigma}^{KM}\{Q_K\bar{Q}_IQ_M\}).
\end{align}
The third term can be dealt with if $F\times G$ is taken to be real $\mathrm{SU}(2)\times\mathrm{SU}(2)$ \cite{Gran:2012mg}.
The solution is then $-a=-b=\tfrac{3}{2}Ac_3$ and (cf. \cite{Gran:2012mg})
\begin{equation}
\bar{H}=-\tfrac{2}{3}a\mathit{\bar{\Sigma}}^{KLM}Q_K\bar{Q}_LQ_M.
\end{equation}
For a fundamental representation their is no solution.
\subsection*{Coupling to supergravity}
In the case of gauge transforming scalars, $\bar{H}_{\mathrm{\textsc{sg}}}$ cannot be solved separately. The two sectors must therefore be added in advance
\begin{align}
\mathit{\Sigma}^{I}\bar{H}\stackrel{!}{=}&a\mathit{\Sigma}^{KL}\{Q_K\bar{Q}_LQ_I\}+b\mathit{\Sigma}^{LM}\{Q_I\bar{Q}_LQ_M\}\nonumber\\
&+\tfrac{1}{4}\lambda\mathit{\Sigma}^{KJ}\mathrm{tr}(Q_{(I}\bar{Q}_{K)})Q_J\nonumber\\
&+\tfrac{1}{4}\lambda\mathit{\Sigma}^{IK}\left(\mathrm{tr}(Q_{(J}\bar{Q}_{K)})Q_J-\tfrac{1}{4}\mathrm{tr}(Q_J\bar{Q}_J)Q_K\right)
\end{align}
\begin{align}
\mathit{\Sigma}^{I}\bar{H}=&A\mathit{\Sigma}^{KL}(\{Q_I\bar{Q}_KQ_L\}+\{Q_K\bar{Q}_LQ_I\}-\{Q_K\bar{Q}_IQ_L\})\nonumber\\
&+\mathit{\Sigma}^{IK}\left(B\{Q_J\bar{Q}_KQ_J\}+C\{Q_K\bar{Q}_JQ_J\}+D\{Q_J\bar{Q}_JQ_K\}\right)\nonumber\\
&+X\lambda\mathit{\Sigma}^{IK}\left(\mathrm{tr}(Q_{(K}\bar{Q}_{J)})Q_J-\tfrac{1}{8}\mathrm{tr}(Q_J\bar{Q}_J)Q_K\right).
\end{align}
This time, the $D$-term is needed to fix the supergravity sector. Still, $\mathrm{SU}(2)\times\mathrm{SU}(2)$ is the only possibility, in which case it can be manipulated as
\begin{align}
\mathit{\Sigma}^{I}\bar{H}\stackrel{!}{=}&-a\mathit{\Sigma}^{KL}Q_K\bar{Q}_LQ_I-b\mathit{\Sigma}^{LM}Q_I\bar{Q}_LQ_M\nonumber\\
&+\tfrac{1}{8}\lambda\mathit{\Sigma}^{KJ}(Q_I\bar{Q}_KQ_J-Q_K\bar{Q}_JQ_I)\nonumber\\
&+\tfrac{1}{4}\lambda\mathit{\Sigma}^{IK}\left(\mathrm{tr}(Q_{K}\bar{Q}_{J})Q_J-\tfrac{1}{4}\mathrm{tr}(Q_J\bar{Q}_J)Q_K\right)
\end{align}
\begin{align}
\mathit{\Sigma}^{I}\bar{H}=&\tfrac{3}{2}Ac_3\mathit{\Sigma}^{KL}(Q_I\bar{Q}_KQ_L+Q_K\bar{Q}_LQ_I)\nonumber\\
&+D\tilde{e}_3\mathit{\Sigma}^{IK}Q_K\mathrm{tr}(\bar{Q}_JQ_J)\nonumber\\
&+X\lambda\mathit{\Sigma}^{IK}\left(\mathrm{tr}(Q_{K}\bar{Q}_{J})Q_J-\tfrac{1}{8}\mathrm{tr}(Q_J\bar{Q}_J)Q_K\right).
\end{align}

One can fix $X=\tfrac{1}{4}$ and $D=-\tfrac{\lambda}{32}$. The coupling constants fulfil $a-b=-\tfrac{\lambda}{4}$ and $a+b=-3Ac_3$ (see also \cite{Gran:2012mg}). Then $\bar{H}=\bar{H}_{\mathrm{\textsc{cs}}}+\bar{H}_{\mathrm{\textsc{sg}}}$ with
\begin{align}
\bar{H}_{\mathrm{\textsc{cs}}}=&-\tfrac{1}{3}(a+b)\mathit{\bar{\Sigma}}^{KLM}Q_K\bar{Q}_LQ_M-\tfrac{\lambda}{32}\mathit{\bar{\Sigma}}^{K}Q_K\mathrm{tr}(Q_J\bar{Q}_J)
\end{align}
and
\begin{align}
\bar{H}_{\mathrm{\textsc{sg}}}=&\tfrac{1}{4}\lambda\mathit{\bar{\Sigma}}^{K}\left(\mathrm{tr}(Q_{K}\bar{Q}_{J})Q_J-\tfrac{1}{8}\mathrm{tr}(Q_J\bar{Q}_J)Q_K\right)+2K\mathit{\bar{\Sigma}}^{K}Q_K.
\end{align}
In order to solve $\bar{H}=0$ for $K$, only one $\mathrm{SO}(8)$ component can be non-zero. This leads again to $|\mu\ell|^{-1}=3$.\\

\noindent One possibility for a fundamental gauge group is $\mathrm{SO}(N)$. In this case,
\begin{align}
\mathit{\Sigma}^{I}\bar{H}=&Dc_1\mathit{\Sigma}^{IK}Q_{\beta}^KQ_{J}^\alpha Q_{\alpha}^J\nonumber\\
&+X\lambda\mathit{\Sigma}^{IK}\left(Q_{K}^\alpha Q_{\alpha}^JQ_{\beta}^J-\tfrac{1}{8}Q_J^\alpha Q_{\alpha}^JQ_{\beta}^K\right)
\end{align}
\begin{align}
\mathit{\Sigma}^{I}\bar{H}\stackrel{!}{=}&2a\mathit{\Sigma}^{KL}Q_K^\alpha Q^L_\beta Q^I_\alpha\nonumber\\
&+\tfrac{1}{4}\lambda\mathit{\Sigma}^{KJ}Q_{I}^\alpha Q^{K}_\alpha Q^J_\beta+\tfrac{1}{4}\lambda\mathit{\Sigma}^{IK}\left(Q_{J}^\alpha Q^{K}_\alpha Q^J_\beta-\tfrac{1}{4}Q_J^\alpha Q^J_\alpha Q^K_\beta\right).
\end{align}

The conditions are $a=-\tfrac{\lambda}{8}$ and $Dc_1-\tfrac{X}{8}\lambda=-\tfrac{1}{16}\lambda$. A choice is $X=\tfrac{1}{4}$ and $Dc_1=-\tfrac{1}{32}\lambda$. Then
\begin{align}
\bar{H}=&\tfrac{\lambda}{16}\mathit{\bar{\Sigma}}^{K}(4Q_{K}^\alpha Q_{\alpha}^JQ_{\beta}^J-Q_J^\alpha Q_{\alpha}^JQ_{\beta}^K)+2K\mathit{\bar{\Sigma}}^{K}Q^K_\beta.
\end{align}
This formula can be used to obtain different values for $\mu\ell$ depending on how many of the $\mathrm{SO}(N)$ components are chosen to be non-zero. Following \cite{Nilsson:2013fya}, one can take them as $Q^I_\alpha=v\updelta^I_\alpha$ where $\alpha=1,...,p\leq 8$. One arrives at the formula (agreeing with \cite{Nilsson:2013fya})
\begin{equation}
|\mu\ell|^{-1}=\big|\tfrac{4}{p}-1\big|.
\end{equation}

\end{document}